\documentclass[runningheads]{llncs}
\usepackage[T1]{fontenc}
\usepackage{graphicx}
\usepackage{color}
\usepackage{amsmath,amssymb,amsfonts,mathtools}
\usepackage{booktabs}
\usepackage{hyperref}
\hypersetup{colorlinks=true,linkcolor=[rgb]{0.1,0.2,0.55},citecolor=[rgb]{0.1,0.2,0.55},urlcolor=[rgb]{0.1,0.2,0.55}}

\hypersetup{pdfauthor={Mikhail Belokon, Rostislav Berezovskiy},
            pdftitle={A game-theoretic analysis of the market microstructure in decentralized exchanges with concentrated liquidity},
            pdfsubject={Mean field games; concentrated liquidity; decentralized exchanges},
            pdfkeywords={decentralized finance, concentrated liquidity, mean field games, liquidity provision, stochastic control}}
\spnewtheorem{assumption}{Assumption}{\bfseries}{\itshape}
\makeatletter
\@ifundefined{proof}%
 {\newenvironment{proof}[1][Proof]{\par\addvspace{\medskipamount}\noindent\textit{#1.}\enspace\ignorespaces}{\par\addvspace{\medskipamount}}}%
 {\renewenvironment{proof}[1][Proof]{\par\addvspace{\medskipamount}\noindent\textit{#1.}\enspace\ignorespaces}{\par\addvspace{\medskipamount}}}
\makeatother
\newcommand{\R}{\mathbb{R}}
\newcommand{\E}{\mathbb{E}}
\newcommand{\Prob}{\mathbb{P}}
\newcommand{\norm}[1]{\left\lVert#1\right\rVert}
\newcommand{\sD}{\sigma_h}                  
\newcommand{\leff}{\lambda_{\mathrm{eff}}}  
\newcommand{\ILe}{\mathrm{IL}}              
\newcommand{\kz}{k_0}                       
\newcommand{\Rfee}{R_{\mathrm{fee}}}        
\newcommand{\Rcyc}{R_{\mathrm{cyc}}}        
\newcommand{\Vbar}{\bar{\mathcal{V}}}       
\newcommand{\Lbar}{\bar{L}}                 
\begin{document}
\title{A game-theoretic analysis of the market microstructure in decentralized exchanges with concentrated liquidity}
\titlerunning{Market microstructure in concentrated-liquidity DEXs}
\author{Mikhail~Belokon
\and Rostislav~Berezovskiy}

\author{Mikhail~Belokon\inst{1}\thanks{Corresponding author.} \and 
Rostislav~Berezovskiy\inst{1}}
%
%

\institute{Vega Institute Foundation, Moscow, Russia}

\institute{Vega Institute Foundation, Moscow, Russia
}
\maketitle
{
\renewcommand{\thefootnote}{}
\footnotetext{%
  \textit{E‑mail addresses:} belokonmike@gmail.com$^*$ (M.~Belokon),
  rostislavberezovskiy@gmail.com (R.~Berezovskiy)/
}
}

\begin{abstract}
We study strategic liquidity provision in decentralized exchanges with
concentrated liquidity as a mean-field game. Liquidity providers choose the
width of their price range and when to rebalance it, pay a transaction cost at
each rebalancing, and share the fees collected at each tick in proportion to
their liquidity, so that each provider's income depends on the state of the
population. We formulate the problem as a controlled continuous-time Markov chain
on a finite state space. In an averaged version of the game, in which a rebalanced
position is re-centered at a draw from the long-run law of the price, a mean-field
equilibrium exists. Because each provider's share of the fees falls as competing
liquidity is added, the equilibrium liquidity profile is unique. For the exact
dynamics, with rebalancing at the range boundary and fast execution, we obtain
closed forms for the optimal width, the value function and the stationary
law of a position, together with a critical volatility above which
provision does not pay. Free entry holds a pool strictly below this threshold, at
a margin set by the rebalancing cost.

The critical volatility is computed from the fee tier, traded volume,
active-tick liquidity and price alone, with no fitted parameter. On two and a
half years of hourly data from five WETH/USDC pools on Ethereum and Base, it
tracks realized volatility. On the Ethereum $0.3\%$ pool the ratio of realized to critical volatility is less than one, and optimal value $w^*$ exists. On the standard Uniswap fee tiers the critical volatility coincides with the on-chain
implied volatility of the pool.

\end{abstract}
%
\section{Introduction}\label{sec:intro}

\emph{Decentralized exchanges} (DEXs) allow users to swap assets without a broker or a traditional order book \cite{uniswap,xu2023}. Many DEXs use \emph{liquidity pools}, where users supply assets to an \emph{automated market maker} (AMM) that determines execution prices from the state of the pool according to a specified pricing rule or invariant. The reserves are supplied by \emph{liquidity providers} (LPs), who receive a share of the fees generated by trades. This income comes with market risk. Arbitrage keeps the pool price aligned with the external market price, and as that price moves, the composition of an LP's position changes. As a result, the value of the position can fall below that of simply holding the same assets outside the pool. This difference is commonly referred to as \emph{impermanent loss} (IL) \cite{xu2023,loesch2021}.

Uniswap~V3 \cite{uniswap} introduced \emph{concentrated liquidity}, which is the setting of this work. In V2, a provider's liquidity is spread over the entire price domain, whereas V3 allows the provider to choose an interval $[a,b]$ in which the liquidity is active. While the price remains in this interval, the position earns trading fees. Once the price leaves the interval, the position becomes inactive and consists of only one of the two assets until the provider takes action. The price domain is discretized into \emph{ticks}, the nodes of a logarithmic grid
$t_0 < \dots < t_M$ with uniform step
$h := \ln(t_m/t_{m-1})$. For example, $h = 60\ln 1.0001 \approx 0.006$ for the $0.3\%$ fee tier and $h = 10\ln 1.0001 \approx 0.001$ for the $0.05\%$ tier. A narrower range concentrates more liquidity around the current price and therefore gives the position a larger fee share while it is active \cite{uniswap,heimbach2022}. At the same time, a narrow range stops earning fees once the price leaves it, and the rate of impermanent loss increases as the range becomes narrower \cite{tang,mmr}. The resulting problem is to choose the range and determine when to move it as the price changes, while accounting for competition between providers and the cost of rebalancing. We model the external price as a \emph{geometric Brownian motion} (GBM) with volatility $\sigma$, take order flow to be exogenous, and assume that each rebalancing costs $g > 0$ per unit of capital.

A provider's fee income also depends on the other providers in the pool. Fees are shared among all positions that cover the tick in proportion to their liquidity, so a larger concentration of competing liquidity reduces an individual provider's share. Popular pools may contain thousands of providers, which makes a direct $N$-player formulation difficult to analyze. Mean field game (MFG) theory replaces this large strategic interaction with a representative-player problem in which a provider responds to the population distribution rather than to every individual competitor. In the limit $N \to \infty$, the equilibrium is described by a backward Hamilton--Jacobi--Bellman (HJB) equation and a forward Kolmogorov (KFE) equation. The two are coupled through the consistency condition that the distribution generated by the provider's optimal strategy coincides with the distribution used to evaluate that strategy. We build on Markovian MFGs on finite state spaces in continuous time \cite{kolokoltsov2010book,kolokoltsov2012finite,kolokoltsov_malafeyev2019,kolokoltsov2017hjb}. MFG methods have also been applied to DEXs in \cite{he2024,bayraktar,baggiani}, while equilibrium in a static concentrated-liquidity market has been characterized in closed form in \cite{hasbrouck2025}. To our knowledge, a closed-form treatment of the dynamic problem, in which providers choose both a range and a rebalancing policy while competing for fees, has not been considered before.

The common price creates an additional difficulty for the mean-field formulation. When a position is rebalanced, its center is set by the current market price, which is shared by all providers. Consequently, conditional on a particular price path, the distribution of positions is random and does not admit a deterministic forward equation in the usual sense. We therefore use two related formulations. For the existence and uniqueness results, we study an \emph{averaged} game in which the center of a newly rebalanced position is drawn from the long-run distribution of the price. For the closed-form analysis, we return to the exact rebalancing rule and describe a position by its \emph{offset}, defined as the distance between its center and the current tick. This representation reduces the equilibrium condition to a single dimensionless ratio, namely realized volatility divided by a \emph{critical volatility} $\sigma_c$. The latter is computed from observable pool quantities. Section~\ref{sec:numerics} evaluates this ratio using on-chain data from five WETH/USDC pools on two chains. Section~\ref{sec:related} reviews related work, Section~\ref{sec:problem} introduces the model, Section~\ref{sec:analysis} presents the analysis, and the longer proofs are collected in the appendix.

\section{Related work}\label{sec:related}

\paragraph{Automated market makers and their providers.}
The mathematical study of AMMs was initiated by \cite{angeris2020}, who identified
constant function market makers with invariant $G(x,y)=k$ and convex trading set,
and proved price convergence under arbitrage. The sign of LP profit is governed by
the curvature of $G$ against the order flow \cite{evans2021}, and DEX architectures
are surveyed in \cite{xu2023}. Provider behavior is documented in
\cite{heimbach2021} for Uniswap~V2, and in \cite{heimbach2022} for V3, where the
chosen width is the dominant driver of a position's risk and return and mean
returns on volatile pairs are typically negative. Machine-learned dynamic
strategies earn $13$--$23\%$ more fee income than uniform allocation
\cite{urusov2025backtesting,urusov2025dynamic}.

\paragraph{Impermanent loss and loss-versus-rebalancing.}
Exact IL formulas for concentrated positions are available in
\cite{loesch2021,tang}, and we adopt the liquidity-cost-per-unit form of
\cite{tang} in Section~\ref{sec:problem}. A related viewpoint is
\emph{loss-versus-rebalancing} (LVR) \cite{mmr}, which for a range position is
inversely proportional to its width. The same inverse-width dependence reappears
in the IL rate of our model, and to leading order the two coincide
(Section~\ref{subsec:leff}).

\paragraph{One provider at a time.}
The single-provider problem is one of stochastic control. \cite{cartea2023}
introduce predictable loss for constant function and concentrated-liquidity
markets, and \cite{cartea2024} solve the dynamic range choice of a single
provider in closed form, with ranges that tilt when the exchange rate has a
stochastic drift. Strategic provision in Uniswap~V3 is studied in
\cite{neuder2021}; the mean--variance profile of LP payoffs and the design of pricing functions are studied in \cite{bergault2024}. These models take the
aggregate depth of the pool as given. In ours it is endogenous, since each
provider's income is diluted by the liquidity of its competitors. For a fixed
aggregate our best-response problem is itself a single-provider problem
(Proposition~\ref{prop:wstar}). What the population adds is that the aggregate is
set by entry and exit, and this fixes the participation margin
(Proposition~\ref{prop:margin}) that Section~\ref{sec:numerics} tests.

\paragraph{Providers in equilibrium.}
With competition, \cite{tang} characterize an $N$-player Nash equilibrium, unique
under diagonal strict concavity \cite{rosen}. Equilibrium liquidity in a
\emph{static} concentrated-liquidity market is characterized in closed form by
\cite{hasbrouck2025}, where a provider commits to a range once. In our model the
control is a rebalancing policy, whose cost per cycle shapes the optimal width and
the participation margin.

\paragraph{Mean field games.}
MFGs are applied to DEXs by \cite{bayraktar}, a numerical V3 calibration with
heterogeneous provider types, and by \cite{he2024}, which optimizes the
invariant. Dynamic-fee optimization \cite{baggiani} is also related. We use the
machinery for controlled finite-state Markov chains in continuous time
\cite{kolokoltsov2010book,kolokoltsov2012finite,kolokoltsov_malafeyev2019,kolokoltsov2017hjb}.
General existence results for finite-state MFGs are in
\cite{gomes2013,cecchin2020}.

\paragraph{Where this paper sits.}
The single-provider models are closed-form but take aggregate depth as given, the
static equilibrium models have no rebalancing, and the DEX mean field models are
dynamic but numerical. We carry the problem with competition through to closed
forms, using exact IL, which captures the asymmetry between the two range
boundaries, and we test the resulting participation margin on data.

\section{Problem statement}\label{sec:problem}

This section defines the price grid, a position, its impermanent loss, its
rebalancing rate and its fee income. Table~\ref{tab:notation} collects the
notation.

\begin{table}[t]
\centering
\caption{Recurring notation. The last column points to the definition.}
\label{tab:notation}
\footnotesize
\setlength{\tabcolsep}{4pt}
\begin{tabular}{@{}lp{7.1cm}l@{}}
\toprule
Symbol & Meaning & Where\\
\midrule
$h,\ M$ & log-tick step; ticks are indexed $0,\ldots,M$ & Def.~\ref{def:grid}\\
$S_t,\ \sigma$ & reference price (GBM); its volatility & Ass.~\ref{ass:exo}\\
$m^*(t),\ \sD,\ \pi$ & active tick; tick volatility $\sigma/h$; its uniform law & Ass.~\ref{ass:exo}\\
$f,\ \mathcal{V},\ \Vbar$ & fee rate; order flow; its local level & Ass.~\ref{ass:exo}, \eqref{eq:Lbar}\\
$j,\ w,\ n$ & center; half-width in ticks; offset $m^*(t)-j$ & Def.~\ref{def:pos}, \S\ref{subsec:leff}\\
$\varepsilon_w,\ \ell_w$ & cost of unit liquidity; liquidity per tick & \eqref{eq:eps_centered}\\
$\varphi,\ \eta$ & boundary IL, total and asymmetry & \eqref{eq:phi_eta}\\
$\leff,\ \rho_w$ & rebalancing rate $\sD^2/w^2$; cycle occupation law & Def.~\ref{def:leff}, \eqref{eq:rho}\\
$L_0,\ L(k,\mu),\ \Lbar$ & background, tick and active-tick liquidity & \eqref{eq:aggL}, \eqref{eq:Lbar}\\
$\Rfee,\ \Rcyc$ & unconditional income; cycle income rate & \eqref{eq:fee}, \eqref{eq:Rcyc}\\
$\alpha,\ B_\alpha$ & fee-decay exponent; its constant & Lem.~\ref{lem:RB}, Rem.~\ref{rem:alpha}\\
$g,\ g_0,\ \kappa$ & rebalancing cost per unit capital; its two parts & Def.~\ref{def:cost}, \eqref{eq:gsplit}\\
$\delta$ & rebalancing indicator ($0$ hold, $1$ rebalance) & Def.~\ref{def:control}\\
$\Gamma,\ \chi$ & execution rate; latency ratio $\leff/\Gamma$ & Def.~\ref{def:control}, Ass.~\ref{ass:Gamma}\\
$R_0,\ \Pi$ & cost rate of a cycle; profit rate & \eqref{eq:Pi}\\
$\sigma_c,\ s$ & critical volatility $2\sqrt{B_1h}$; margin $\sigma/\sigma_c$ & \eqref{eq:wstar_c}\\
$s_\star,\ s_{\mathrm{excl}}$ & free-entry margin; exclusion margin & \eqref{eq:sstar}, \eqref{eq:sexcl}\\
\bottomrule
\end{tabular}
\end{table}

\subsection{The tick grid, the players and the exogenous processes}\label{subsec:model_scope}

We work on a fixed \emph{finite interval} $[0, T]$, $T > 0$. The stochastic basis
$(\Omega, \mathcal{F}, \{\mathcal{F}_t\}_{t\ge0}, \Prob)$ satisfies the usual
conditions.

\begin{definition}[Tick grid]\label{def:grid}
The tick grid consists of the prices $t_m = t_0e^{mh}$, $m = 0,\ldots,M$, where the
logarithmic step $h := \ln(t_m/t_{m-1}) > 0$ is fixed by the protocol, as described
in the Introduction. Following Uniswap, we call the node $t_m$ \emph{tick}~$m$ and
say that the price is in tick $m$ when it lies in $[t_m, t_{m+1})$, so the interval
between two adjacent ticks is labelled by its left endpoint.
\end{definition}

\begin{assumption}[exogenous dynamics]\label{ass:exo}
We fix the following objects.
\begin{enumerate}
  \item \textbf{Price.} The reference price $S_t$ follows the
        \emph{geometric Brownian motion}
        \begin{equation}\label{eq:GBM}
          dS_t = \tfrac12\sigma^2 S_t\,dt + \sigma S_t\,dW_t,
          \quad S_0 > 0,\quad \sigma > 0,
        \end{equation}
        with $W_t$ a standard Brownian motion. The drift is chosen so that
        $d(\ln S_t) = \sigma\,dW_t$ by It\^o's formula. The symmetry of exits
        used below holds exactly when the log-price is driftless.
  \item \textbf{Order flow.} The traded volume per unit time between ticks $k-1$
        and $k$, in the numeraire, is a deterministic function
        $\mathcal{V}_t(k) \ge v_{\min} > 0$, independent of the providers' actions,
        and time-independent, $\mathcal{V}_t(k) \equiv \mathcal{V}(k)$, in the
        stationary analysis. The lower bound $v_{\min}$ is used only in the
        proof of uniqueness (Theorem~\ref{thm:unique}), where it pins down the
        liquidity at every tick and not only at the active one.
  \item \textbf{Active tick.} The active tick is
        $m^*(t) := \lfloor \ln(S_t/t_0) / h \rfloor$, the tick whose interval
        holds the current price. As $h \to 0$, $m^*(t)h \to \ln(S_t/t_0)$, and we
        use throughout the
        approximation
        \begin{equation}\label{eq:tick_SDE}
          dm^*(t) \approx \sD\,dW_t,
          \qquad \sD := \sigma / h .
        \end{equation}
        On the grid we realize it as the symmetric nearest-neighbor random walk
        on $\{0,\ldots,M\}$ with rate $\tfrac12\sD^2$ to each neighbor, reflected
        at the two ends. Its stationary law is the uniform law
        $\pi_k = 1/(M+1)$, which is the $\pi$ used below.
  \item \textbf{Players.} There is a \emph{continuum of identical liquidity
        providers}, risk neutral, each with unit capital, so no single provider
        affects the aggregate. Costs and impermanent loss are charged to the
        running payoff and are not deducted from the stake. The process $m^*(t)$
        is \emph{common} to all providers, since there is one price.
\end{enumerate}
\end{assumption}

\subsection{Widths and the provider's position}

\begin{definition}[Admissible widths]\label{def:W}
The admissible half-widths of a position, measured in ticks, form a finite set
$\mathcal{W} := \{0, w_1, \ldots, w_N\}$ with $0 < w_1 < \dots < w_N \le M/2$.
The value $w = 0$ means that the provider is out of the pool and supplies no
liquidity.
\end{definition}

The general theorems use only the finiteness of $\mathcal{W}$. The closed forms of
Section~\ref{subsec:stationary} are the limit of a fine set $\mathcal{W}$, in which
the half-width is treated as continuous.

\begin{definition}[Provider's position]\label{def:pos}
Let $X$ be the \emph{numeraire} asset. Choosing a center
$j \in \{w,\ldots,M-w\}$ (so that the range fits inside the grid) and half-width
$w > 0$, the provider supplies liquidity over the \emph{price range}
$R(j, w) := \{k \in \mathbb{Z} : |k - j| \le w\}$, with price boundaries
$a := t_{j-w}$ and $b := t_{j+w}$. Following the dollar-value expression for a
liquidity position in \cite{tang}, the \emph{cost of supplying one unit of
liquidity} at the price $q \in [a, b]$ is
\begin{equation}
  \varepsilon_{a,b}(q) := 2\sqrt{q} - \sqrt{a} - \frac{q}{\sqrt{b}} .
\end{equation}
For a centered position, $q = t_j = \sqrt{ab}$, this evaluates to
\begin{equation}\label{eq:eps_centered}
  \varepsilon_w := \varepsilon_{a,b}(t_j)
  = 2\sqrt{t_j}\,\bigl(1 - e^{-wh/2}\bigr) \approx \sqrt{t_j}\,wh
  \quad (wh \to 0).
\end{equation}
A provider with unit capital supplies the liquidity $\ell_w := 1/\varepsilon_w$ to
each tick of the chosen range, so its contribution to tick $k$ is
$\ell(k, j, w) := \ell_w\, \mathbf{1}\{k \in R(j, w)\}$.
\end{definition}

Uniswap~V2 is the limit $a \to 0$, $b \to \infty$, in which
$\varepsilon_{a,b}(q) \to 2\sqrt{q}$. As $w$ decreases, $\varepsilon_w$ decreases
in proportion to $wh$, so more liquidity is placed per tick.

\subsection{Impermanent loss}\label{subsec:IL}

Impermanent loss has exact closed forms for concentrated liquidity
\cite{loesch2021,tang}, which we adopt without re-deriving them. Write
$\ILe(j,w,q',q)$ for the IL of a position $(j,w)$, as a fraction of unit
capital, when the price moves from $q$ to $q'$. The exact expression and its
in-range reduction for a centered position are recalled in
Appendix~\ref{app:proofs}, Eqs.~\eqref{eq:ILexact}--\eqref{eq:ILexact_inrange}.
The benchmark is buy-and-hold of the entry composition, so the provider is
assumed to arrive holding the two assets in the proportions a centered range
requires.

A provider who rebalances at the range boundary realizes the IL at a boundary.
At the two boundaries the exact IL of a centered position is
\begin{equation}\label{eq:ILrebal}
  \ILe\bigl|_{q'=b} = \frac{e^{wh} - e^{wh/2}}{2}
  \quad\text{(upper)},
  \qquad
  \ILe\bigl|_{q'=a} = \frac{1 - e^{-wh/2}}{2}
  \quad\text{(lower)},
\end{equation}
with the sum and difference
\begin{align}\label{eq:phi_eta}
  \varphi(w) &:= \ILe\bigl|_{q'=b} + \ILe\bigl|_{q'=a}
             = \bigl(e^{wh/2} - 1\bigr)\cosh\tfrac{wh}{2}, \notag\\
  \eta(w)    &:= \ILe\bigl|_{q'=b} - \ILe\bigl|_{q'=a}
             = \bigl(e^{wh/2} - 1\bigr)\sinh\tfrac{wh}{2}.
\end{align}
Here $\varphi$ is the total boundary IL and $\eta > 0$ its asymmetry. Exiting
through the lower boundary is cheaper. The asymmetry comes from valuing the
position in the numeraire over geometrically placed boundaries, since the
log-price itself is driftless. For narrow ranges
\begin{equation}\label{eq:phi_exp}
  \varphi(w) = \frac{wh}{2} + \frac{(wh)^2}{8} + \frac{(wh)^3}{12} + O\bigl((wh)^4\bigr),
  \qquad
  \eta(w) = \frac{(wh)^2}{4} + O\bigl((wh)^3\bigr),
\end{equation}
so the boundary IL is linear in the width at leading order. This is the classical
IL approximation, quadratic in the price move, evaluated at a boundary
displacement proportional to $wh$.

\subsection{Effective rebalancing rate and the cycle occupation law}\label{subsec:leff}

A position is revised when the price leaves its range, so the time scale of the
problem is the exit time of the tick process. The \emph{offset} of a position with
center $j$ is $n := m^*(t) - j$, the signed distance in ticks from the center to
the active tick. A position re-centered at the price has $n = 0$, and the stretch
of time from a re-centering to the first time $\tau_{\mathrm{exit}}$ at which
$|n| = w$ is a \emph{rebalancing cycle}. For a Brownian motion with volatility
$\sD$ started at the center, $\E[\tau_{\mathrm{exit}}] = w^2/\sD^2$
\cite{karatzas1998}, and Lemma~\ref{lem:occupation} below shows that the same
identity holds exactly for the lattice walk of Assumption~\ref{ass:exo}.

This boundary-triggered rebalancing is a $\tau$-reset strategy in the sense of
\cite{neuder2021}: the position is re-centered only when the price leaves its band,
and the half-width of the band, $n_\tau$ in their notation, is our $w$. Equivalently,
the offset is reset to $0$ when $|n|$ reaches $w$.

\begin{definition}[Effective rebalancing rate]\label{def:leff}
For half-width $w > 0$ and tick volatility $\sD = \sigma/h$, the rate at which
the price forces a provider to revise its position is
$\leff := \sD^2/w^2 = 1/\E[\tau_{\mathrm{exit}}]$.
\end{definition}

\begin{lemma}[Occupation law over a cycle]\label{lem:occupation}
For the offset walk with generator $\tfrac12\sD^2\Delta_d$, where \(\Delta_d\) is the discrete Laplacian, \(\Delta_d f(n) := f(n+1)-2f(n)+f(n-1)\), started at $n = 0$ and
absorbed at $|n| = w$, the expected time spent at offset $n$ before absorption is
\begin{equation}\label{eq:Green}
  G_w(n) = \frac{w - |n|}{\sD^2}, \qquad |n| \le w,
  \qquad \sum_{n=-w}^{w} G_w(n) = \frac{w^2}{\sD^2} = \E[\tau_{\mathrm{exit}}].
\end{equation}
Consequently the normalized \emph{cycle occupation law}
\begin{equation}\label{eq:rho}
  \rho_w(n) := \leff\, G_w(n) = \frac{w - |n|}{w^2},
  \qquad n \in \{-w, \ldots, w\},
\end{equation}
is a probability vector, a triangular profile centered at $n = 0$.
\end{lemma}
\begin{proof}
See Appendix~\ref{app:proofs}.
\qed
\end{proof}

The product of $\leff$ and the expected boundary IL per cycle,
$\tfrac12\varphi(w)$ (exit is equally likely at either boundary for the driftless
offset), is the impermanent-loss rate
\begin{equation}\label{eq:ILrate}
  \tfrac12\,\leff\, \varphi(w)
  = \frac{\sD^2}{2w^2}\,\bigl(e^{wh/2}-1\bigr)\cosh\tfrac{wh}{2}
  \approx \frac{\sigma^2}{4wh} \quad (wh \to 0).
\end{equation}

To leading order this rate is the loss-versus-rebalancing (LVR) of \cite{mmr}
for a range position. Write $x := wh/2$, so that the range is $[t_je^{-2x},\,t_je^{2x}]$ and
the leading term of~\eqref{eq:ILrate} is $\sigma^2/(8x)$. For a constant-product
pool LVR is $\sigma^2/8$ per unit of value and unit time. The same computation for
a range position, whose risky reserves are $L(q^{-1/2} - b^{-1/2})$ at the
price~$q$, gives $\sigma^2 L\sqrt q/4$, which at the center is
\begin{equation}\label{eq:lvr}
  \frac{\sigma^2}{8\,(1-e^{-x})} = \frac{\sigma^2}{8x}\Bigl(1 + \frac{x}{2} + O(x^2)\Bigr)
\end{equation}
per unit of value, and~\eqref{eq:ILrate} agrees with it in its first two terms.
Concentration thus multiplies the arbitrage loss of a full-range position by about
$1/x$, and a narrow range is a leveraged short-volatility position.

\subsection{Fee-income model}\label{subsec:fee}

Two fee expectations appear in the model and must not be conflated, the
\emph{unconditional} income of a static position, averaged over the price law on
the whole grid, and the \emph{cycle rate} of a provider who rebalances at the
boundary and is therefore in range throughout each cycle.

\begin{definition}[Aggregate liquidity and unconditional fee income]\label{def:fee}
Let $L_0 > 0$ be an exogenous background liquidity present in every tick. The aggregate liquidity in tick $k$ induced by the position
distribution $\mu$ is
\begin{equation}\label{eq:aggL}
  L(k,\mu) := L_0 + \sum_{(j',w'):\,w'>0} \mu(j',w')\, \ell(k,j',w') \;\ge\; L_0 .
\end{equation}
For a provider with position $(j,w)$, $w>0$, under the measure $\mu$, the
\emph{unconditional} expected fee income is
\begin{equation}\label{eq:fee}
  \Rfee(j,w,\mu)
  := \sum_{k \in R(j,w)} \pi_k \, f \,\mathcal{V}(k)\,
  \frac{\ell(k,j,w)}{L(k,\mu)},
\end{equation}
and $\Rfee(j,0,\mu) := 0$. Here $f \in (0,1)$ is the protocol fee rate and $\pi$ is
the stationary law of the active tick (Assumption~\ref{ass:exo}).
\end{definition}

\begin{definition}[Cycle income rate]\label{def:Rcyc}
A provider that rebalances at the range boundary is in range throughout each cycle,
and at offset $n$ it earns its share of the fee flow at the active tick $j+n$.
Weighting by the cycle occupation law of Lemma~\ref{lem:occupation}, its
\emph{cycle income rate}, the expected income per cycle divided by the expected
cycle length, is
\begin{equation}\label{eq:Rcyc}
  \Rcyc(j,w,\mu)
  := \sum_{n=-w}^{w} \rho_w(n)\, f\,\mathcal{V}(j+n)\,
    \frac{\ell(j+n,j,w)}{L(j+n,\mu)} .
\end{equation}
\end{definition}

The two functionals play different roles. The unconditional $\Rfee$ is the income
of the averaged game of Section~\ref{subsec:layers}, and every break-even and
optimal-width formula of Section~\ref{subsec:stationary} involves $\Rcyc$. Write
$P_{\mathrm{in}}(w) := \sum_{k\in R(j,w)}\pi_k$ for the share of time the price
spends in a given range, which is about $(2w+1)/(M+1)$. When volume and liquidity
are locally flat, $\Rcyc \approx \Rfee/P_{\mathrm{in}}(w)$, so the unconditional
income understates the income of a rebalancing provider by a factor of order
$M/w$. Both functionals are positive-weighted sums of the shares $\ell/L$, so each
falls when liquidity is added inside the range of the position.

When every active provider rebalances at its range boundary, every active position
covers the active tick at all times, up to execution latency, and the liquidity at
the active tick is the deterministic stack
\begin{equation}\label{eq:Lbar}
  \Lbar := L_0 + \sum_{w'>0} \theta(w')\,\ell_{w'},
\end{equation}
where $\theta(w')$ is the mass of active providers of half-width $w'$. With volume
locally flat at the level $\Vbar$ over a window of width $2w$, the cycle rate is
then $\Rcyc = f\Vbar\,\ell_w/\Lbar = f\Vbar/(\Lbar\,\varepsilon_w)$.

\begin{lemma}[Cycle income rate]\label{lem:RB}
Let every active provider rebalance at its range boundary without delay, and let
volume be locally flat at the level $\Vbar$ over the range of the position. Then the
cycle income rate of a provider with half-width $w$ is
\begin{equation}\label{eq:B1}
  \Rcyc(w) = \frac{B_1}{w}\bigl(1 + O(wh)\bigr),
  \qquad
  B_1 = \frac{f\Vbar}{\sqrt{\bar q}\,h\,\Lbar},
\end{equation}
where $\bar q$ is the local price level and $\Lbar$ is the active-tick
liquidity~\eqref{eq:Lbar}.
\end{lemma}
\begin{proof}
Every active position covers the active tick, so the liquidity there is $\Lbar$
throughout the cycle, and~\eqref{eq:Rcyc} reduces to
$\Rcyc = f\Vbar\,\ell_w/\Lbar = f\Vbar/(\Lbar\,\varepsilon_w)$.
By~\eqref{eq:eps_centered}, $\varepsilon_w = \sqrt{\bar q}\,wh\,\bigl(1 + O(wh)\bigr)$,
which gives~\eqref{eq:B1}.
\qed
\end{proof}

A wider position spreads the same unit of capital over more ticks, so its liquidity
per tick, and with it its share of the fee flow at the active tick, falls as $1/w$.
The constant $B_1$ is set by the fee rate $f$, the local order flow $\Vbar$ and the
liquidity $\Lbar$, which contains the background liquidity $L_0$ and the liquidity of
competitors. An individual provider takes $\Lbar$, and hence $B_1$, as given, since
its own liquidity is negligible, and free entry adjusts $\Lbar$ through the mass of
active providers (Section~\ref{subsec:stationary}). Theorems~\ref{thm:exist}
and~\ref{thm:unique} do not use the lemma.

\begin{remark}\label{rem:alpha}
The closed forms of Section~\ref{subsec:stationary} are stated for the reduced form
$\Rcyc = B_\alpha/w^\alpha$ with $\alpha \in (0,1]$ and $B_\alpha > 0$, of
which~\eqref{eq:B1} is the case $\alpha = 1$. An exponent $\alpha < 1$ allows a
flatter dependence on the width when volume varies across the ticks of the range. It
is not derived from the model, and every numerical result in
Section~\ref{sec:numerics} uses $\alpha = 1$ with $B_1$ from~\eqref{eq:B1}.
\end{remark}

\subsection{State space, control and costs}\label{subsec:state_control}

\begin{definition}[State space]\label{def:state}
The state of an individual provider is the pair $(j, w)$ of the central tick and
the half-width of its position. The state space is
$\mathcal{S} := \{0, 1, \ldots, M\} \times \mathcal{W}$.
\end{definition}

\begin{definition}[Strict and mixed controls]\label{def:control}
A \emph{strict control} is an adapted process \(u_t = (\delta_t, w'_t)\) taking values in
\(\mathcal{U} := \{0,1\} \times \mathcal{W}\). The value \(\delta_t = 0\) means holding
the current position, while \(\delta_t = 1\) means rebalancing to the position
\((m^*(t), w'_t)\).

A \emph{mixed control} randomizes over the strict controls: at each time \(t\)
and in each state \((j,w)\) it prescribes a probability distribution
\(\nu_t(j,w) \in \mathcal{P}(\mathcal{U})\), where \(\mathcal{P}(\mathcal{U})\) is
the simplex of probability distributions on the finite set \(\mathcal{U}\). A strict
control is the special case in which \(\nu_t(j,w)\) puts probability one on a single
action.

When \(\delta_t = 1\), rebalancing opportunities arrive according to a Poisson
process with intensity \(\Gamma > 0\), modeling the delay until the transaction
is included in a block.
\end{definition}

The mixed control is needed only for the fixed-point theorem in the existence proof.
The stationary equilibrium in Section~\ref{subsec:equilibrium} uses a strict
control given by the boundary rule.

\begin{definition}[Transition cost]\label{def:cost}
The cost of moving from state \((j,w)\) to state \((k,w') \neq (j,w)\) is
\(c(j,w,k,w') = \ILe(j,w,t_k,t_j)+g\) if \(w>0\), \(g\) if \(w=0\) and \(w'>0\), and \(0\) if \(w=w'=0\). 

We also set \(c(j,w,j,w)=0\), since a null rebalance executes no transaction. Here \(g>0\) is the transaction cost as a fraction of unit capital, and the IL term is evaluated at the new center \(t_k\), corresponding to the price at re-centering.
\end{definition}

The cost $g$ has two parts of different kinds. Execution carries a \emph{fixed}
charge $g_0$ (gas) independent of position size. Restoring the range carries a
\emph{proportional} charge, because a position that has left its range is held
entirely in one asset, and restoring the range requires converting about half of
it at the venue fee plus slippage. For a provider of capital $c$, per unit of
capital,
\begin{equation}\label{eq:gsplit}
  g \;=\; \frac{g_0}{c} \;+\; \kappa,
  \qquad \kappa \;\approx\; \tfrac12\bigl(f_{\mathrm{ex}} + \varsigma\bigr),
\end{equation}
with $f_{\mathrm{ex}}$ the fee of the venue used for the conversion and $\varsigma$
a slippage allowance. Only the first term falls with capital, so $g$ decreases to
$\kappa$ as $c$ grows.  Providers of different capital therefore differ, in this model, only through $g$.

\subsection{Running payoff and the optimization problem}\label{subsec:utility}

\begin{definition}[Running payoff]\label{def:payoff}
For a provider in state $(j, w)$ with active tick $m^*(t)$ and measure $\mu$, set
\begin{equation}
  r(j, w, \mu, m^*) :=
  \begin{cases}
    \dfrac{\ell(m^*,j,w)}{L(m^*,\mu)}\,f\,\mathcal{V}(m^*),
    & w > 0,\;\; m^*(t) \in R(j, w),\\[6pt]
    0, & \text{otherwise.}
  \end{cases}
\end{equation}
This is the provider's liquidity share in the current tick times the fee flow
there. Its average over $\pi$ is $\Rfee$, and its average over a rebalancing cycle
started at $m^*(0) = j$ is $\Rcyc$.
\end{definition}

\begin{definition}[Terminal condition]\label{def:terminal}
At time $T$ a provider with position $(j, w)$, $w > 0$, incurs the impermanent
loss at the current price, $V_T(j, w) := -\ILe(j, w, t_{m^*(T)}, t_j)$, and
$V_T(j, 0) := 0$.
\end{definition}

The representative provider maximizes expected total profit over $[0, T]$,
\begin{equation}\label{eq:obj}
  \max_{\{u_t\}} \E\!\left[
    \int_0^T r(j_t, w_t, \mu_t, m^*(t))\,dt
    - \sum_{\tau} c(j_{\tau^-}, w_{\tau^-}, j_\tau, w_\tau)
    + V_T(j_T, w_T)
  \right],
\end{equation}
the maximum being over admissible adapted controls and the sum over rebalancing
times. A \emph{mean field equilibrium} (MFE) is a population law that is
reproduced by the optimal control it induces (Definition~\ref{def:MFE}).

\section{Construction and analysis of the solution}\label{sec:analysis}

We work in the standard framework of mean field games for controlled Markov
chains on a finite state space \cite{kolokoltsov2010book,kolokoltsov2012finite,kolokoltsov_malafeyev2019}.
A controlled generator encodes how rebalancing moves the position, the value
function solves a backward HJB equation, the population law evolves by a forward
KFE, and an equilibrium is a fixed point of the map that sends a candidate
population law to the law generated by the best response to it.
Sections~\ref{subsec:layers}--\ref{subsec:uniqueness} set up the averaged game and
prove existence and uniqueness.
Section~\ref{subsec:stationary} returns to the exact
re-centering rule and derives the stationary equilibrium in closed form.

\subsection{Exact and averaged re-centering}\label{subsec:layers}

Under Definition~\ref{def:control} a rebalancing re-centers the position at the
current active tick $m^*(t)$. Since $m^*(t)$ is
common to all providers, the positions of the population move with the price,
their empirical distribution is random given the price path, and no deterministic
forward equation holds for it. The general theorems are therefore proved for an
averaged version of the game.

\begin{assumption}[averaged re-centering]\label{ass:annealed}
In the averaged game a rebalanced position is re-centered at a tick $k$ drawn from
$\pi$, independently across providers and of the past. The running income of a
position $(j,w)$ is its $\pi$-average $\Rfee(j,w,\mu)$ of~\eqref{eq:fee}, and the
impermanent loss charged at a rebalancing and at the terminal time is evaluated at
the drawn tick.
\end{assumption}

The two rules share the long-run law of the center of a tracked position, up to
boundary effects. Under the averaged rule it is $\pi$
by construction. Under the exact rule the center of a position is
$j = m^*(t) - n$, with $n$ its offset. Away from the ends of the grid the dynamics
of the pair $(m^*, n)$ are invariant under translations of the price, and the
stationary law of $m^*$ is uniform, so the stationary law of the pair is the
product of the uniform law with the law of $n$, and $j$ is uniform as well. The
reflecting ends change this only within $w$ ticks of the ends of the grid. This is
a statement about one position averaged over price paths. Given the price, the
population under the exact rule is concentrated within $w$ ticks of it, which is
what the aggregate $\Lbar$ of~\eqref{eq:Lbar} records.

The rules differ in the joint law of price and center, and both fees and
impermanent loss depend on price and center only through their difference. Under
the exact rule a fresh position always covers the active tick. Under the averaged
rule it does so with probability about $(2w+1)/(M+1)$. This is why the income of
the averaged game is $\Rfee$ while that of the exact dynamics is $\Rcyc$, and why
the closed forms of Sections~\ref{subsec:stationary}--\ref{subsec:equilibrium}
are derived for the exact rule, in the offset coordinate, where the price drops
out of the state and rebalancing is the reset $n \mapsto 0$. The averaged game
supplies the general existence and uniqueness theory. The exact dynamics supply
every number that is compared with data. The two are different games, and we
prove no limit theorem that connects them.

\subsection{The HJB--KFE system}\label{subsec:MFG_system}

\begin{definition}[Controlled generator]\label{def:Qmat}
For a strict control $u = (\delta, w') \in \mathcal{U}$,
\begin{align}
  Q^u_{(j,w),(k,\tilde w)} &:= \delta\Gamma\,\pi_k\,
  \mathbf{1}\{\tilde w = w'\}\,\mathbf{1}\{(k,\tilde w) \ne (j,w)\},
  \label{eq:Qoff}\\
  Q^u_{(j,w),(j,w)} &:= -\delta\Gamma(1 - \pi_j\,\mathbf{1}\{w = w'\}).
  \label{eq:Qdiag}
\end{align}
For a mixed control $\nu \in \mathcal{P}(\mathcal{U})$,
$Q^\nu_{(j,w),(k,\tilde w)} := \sum_{u \in \mathcal{U}} \nu(u)\,Q^u_{(j,w),(k,\tilde w)}$,
which is linear in~$\nu$.
\end{definition}

On rebalancing the new center $k$ is drawn from $\pi$ and the new width is the
chosen $w'$. The off-diagonal entries are nonnegative, and each row sums to zero,
since the off-diagonal sum is $\delta\Gamma(1-\pi_j)$ for $w'=w$ and $\delta\Gamma$
for $w'\ne w$.

\begin{definition}[Hamilton--Jacobi--Bellman equation]\label{def:HJB}
The value function $V(t,j,w)$ of the averaged game, the maximal expected profit
over $[t,T]$ of a provider in state $(j,w)$ at time $t$, satisfies
\begin{equation}\label{eq:HJB}
\frac{d}{dt}V(t,j,w) + \Rfee(j,w,\mu_t)
+ \max_{(\delta,w')}
\delta\Gamma\Bigl[\sum_k \pi_k\bigl(V(t,k,w') - c(j,w,k,w')\bigr)
- V(t,j,w)\Bigr]
= 0,
\end{equation}
with the terminal condition
$V(T,j,w) = -\sum_k \pi_k \ILe(j,w,t_k,t_j)$ for $w > 0$ and $V(T,j,0) = 0$, the
$\pi$-average of Definition~\ref{def:terminal}.
\end{definition}

Equation~\eqref{eq:HJB} is the dynamic programming principle over a short interval
$[t, t+dt]$. The provider earns $\Rfee\,dt$, and with probability $\delta\Gamma\,dt$
it rebalances to a center drawn from $\pi$ at the cost $c$, and otherwise it
holds \cite{kolokoltsov2010book,kolokoltsov2017hjb}. The running income does not
depend on the control, so only the rebalancing decision is optimized. The price
does not appear in~\eqref{eq:HJB}, because averaging over the draw of the price
removes it from the state. The value of the exact dynamics carries in addition the
tick generator and the running payoff $r$, and its stationary version is solved in
Theorem~\ref{thm:Vstar}.

The maximand in~\eqref{eq:HJB} is $\delta\Gamma\,\Psi(t,j,w,w')$, where
\[
  \Psi(t,j,w,w') := \sum_k \pi_k\bigl(V(t,k,w') - c(j,w,k,w')\bigr) - V(t,j,w)
\]
is the gain from rebalancing to the half-width $w'$. Thus, rebalancing occurs
when the gain is positive. In that case, the provider chooses
$w'^* \in \arg\max_{w'}\Psi$. Otherwise, the current position is kept
unchanged.

\begin{definition}[Forward Kolmogorov equation]\label{def:KFE}
For a mixed feedback control $\nu_t(j,w) \in \mathcal{P}(\mathcal{U})$,
\begin{equation}\label{eq:KFE}
  \frac{d}{dt}\mu_t(k,\tilde w)
  = \sum_{(j,w) \in \mathcal{S}} \mu_t(j,w)\,
    Q^{\nu_t(j,w)}_{(j,w),(k,\tilde w)}, \quad (k,\tilde w) \in \mathcal{S}.
\end{equation}
\end{definition}

Here $\nu_t(j,w)$ is the mixed control used in the state $(j,w)$, and
$(k,\tilde w)$ is the state reached.

\begin{lemma}[Simplex invariance]\label{lem:simplex}
For every control measurable in~$t$, whether strict or mixed, if
$\mu_0 \in \Sigma_{|\mathcal{S}|}$ then the solution of~\eqref{eq:KFE} satisfies
$\mu_t \in \Sigma_{|\mathcal{S}|}$ for all $t \in [0,T]$.
\end{lemma}
\begin{proof}
See Appendix~\ref{app:proofs}.
\qed
\end{proof}

\begin{definition}[Mean field equilibrium, MFE]\label{def:MFE}
A triple $(\hat{V}, \hat\mu, \hat{\nu})$ is an MFE of the averaged game when three
conditions hold. First, $\hat{V}$ solves~\eqref{eq:HJB} at $\mu_t = \hat\mu_t$.
Second, $\hat\mu$ solves~\eqref{eq:KFE} under the control $\hat{\nu}$. Third,
$\hat{\nu}_t(j,w)$ is supported on the maximizers in~\eqref{eq:HJB} at $\hat\mu$
for a.e.\ $t$.
\end{definition}

\subsection{Existence of the MFE}\label{subsec:existence}

\begin{theorem}[Existence of the MFE]\label{thm:exist}
Under Assumptions~\ref{ass:exo} and~\ref{ass:annealed}, with $\Gamma > 0$ and
$L_0 > 0$, the averaged game has at least one MFE in the class of mixed controls
(Definition~\ref{def:control}).
\end{theorem}
\begin{proof}
See Appendix~\ref{app:proofs}.
\qed
\end{proof}

This is the natural extension of Nash's existence theorem for finite games in mixed strategies to a continuum of players.

The hypothesis $L_0 > 0$ is what makes the income continuous on the simplex, since
$L(k,\mu) \ge L_0$ bounds every share. With it the averaged game also falls within
the general existence theory for finite-state mean field games with mixed
controls (relaxed controls in the terminology of \cite{cecchin2020}). The proof in the appendix is self-contained.

\subsection{Uniqueness of the MFE}\label{subsec:uniqueness}

The fee income depends on the total liquidity at each tick: as more
liquidity accumulates at a tick, each provider receives a smaller share
of the fees. For any two population distributions \(\mu\) and
\(\tilde\mu\), we have
\[
\sum_{(j,w)} \ell(k,j,w)\,[\mu-\tilde\mu](j,w)
= L(k,\mu)-L(k,\tilde\mu)
\]
for every tick \(k\). Therefore, the fee income~\eqref{eq:fee} satisfies
\begin{equation}\label{eq:monotone}
\begin{aligned}
&\sum_{(j,w)}
\bigl[\Rfee(j,w,\mu)-\Rfee(j,w,\tilde\mu)\bigr]
\bigl[\mu(j,w)-\tilde\mu(j,w)\bigr]\\
&\qquad =
-\sum_k \pi_k f\mathcal{V}(k)
\frac{\bigl(L(k,\mu)-L(k,\tilde\mu)\bigr)^2}
{L(k,\mu)L(k,\tilde\mu)}
\le 0,
\end{aligned}
\end{equation}
with equality if and only if \(L(k,\mu)=L(k,\tilde\mu)\) for every
tick \(k\). This inequality expresses the congestion effect in fee
income. It is the monotonicity condition of Lasry and
Lions~\cite{lasrylions2007}, the population-level counterpart of Rosen's
diagonal strict concavity~\cite{rosen}. The condition is
formulated in terms of total liquidity at each tick, rather than the
full population distributions.

\begin{theorem}[Uniqueness of the MFE]\label{thm:unique}
Under the assumptions of Theorem~\ref{thm:exist}, two MFEs of the averaged game
with the same initial law induce the same aggregate liquidity $L(k,\mu_t)$ at every
tick and every time, and the same value function. If, in addition, the maximizer
in~\eqref{eq:HJB} is unique at a.e.\ time and every state, then the equilibrium law
$\mu$ is unique as well.
\end{theorem}
\begin{proof}
See Appendix~\ref{app:proofs}.
\qed
\end{proof}

Without the additional hypothesis two equilibria can differ only through ties in
the rebalancing decision, which leave the payoff-relevant objects unchanged. In
this sense the MFE is unique. Neither theorem uses Lemma~\ref{lem:RB} or the
reduced form of Remark~\ref{rem:alpha}.

Neither theorem needs identical providers either. Let providers belong to finitely
many types $\theta$ that differ in the cost $g_\theta$, as providers of different
capital do by~\eqref{eq:gsplit}, and let $\mu(j,w,\theta)$ be the capital of type
$\theta$ in the state $(j,w)$. No transition changes the type, so the state space
stays finite and each type keeps its mass. Income per unit of capital is
$\Rfee(j,w,\mu)$ for every type and depends on $\mu$ only through the aggregates
$L(k,\cdot)$. The proofs of Theorems~\ref{thm:exist} and~\ref{thm:unique} then
apply without change, on the product of the type-wise simplices and with the
identity~\eqref{eq:monotone} applied to the marginal of $\mu$ over types.
Corollary~\ref{cor:entry} shows how such types sort in the stationary equilibrium.

\subsection{Stationary analysis of the exact dynamics}\label{subsec:stationary}

From here on we return to the exact rule and describe a position by its offset
$n = m^*(t) - j$. Volume is time-independent (Assumption~\ref{ass:exo}), the horizon
is long ($T \gg 1/\leff$), the criterion is the long-run profit per unit of time,
and two further assumptions fix the regime and the class of policies.

\begin{assumption}[fast rebalancing]\label{ass:Gamma}
The execution rate satisfies $\Gamma \gg \leff$. The statements below are the
limits as $\chi := \leff/\Gamma \to 0$.
\end{assumption}

The assumption is not restrictive for the pools of Section~\ref{sec:numerics}.
A rebalancing waits for block inclusion, about $12$ seconds on Ethereum and a few
seconds on a rollup, while a cycle lasts $1/\leff = w^2h^2/\sigma^2$, from hours
for narrow ranges to $25$ to $70$ days at the free-entry widths of
Section~\ref{sec:numerics}. Hence $\chi$ lies between about $10^{-6}$ and
$10^{-3}$. Each cycle is lengthened by a wait of mean
$1/\Gamma$, so the rate of cycles becomes $(1/\leff + 1/\Gamma)^{-1} =
\leff(1 - \chi + O(\chi^2))$, and the income and cost rates below change by a
relative amount of order $\chi$.

\begin{assumption}[boundary-triggered provision]\label{ass:boundary}
We restrict attention to boundary-triggered strategies. An active provider of
half-width $w$ holds its position while the price is inside the range and
rebalances when $|n|$ reaches $w$, re-centering at the current tick with the same
half-width. Thus, each provider chooses a half-width and follows this
boundary-triggered rebalancing rule, or chooses to stay out of the pool.
\end{assumption}

The restriction binds near the margin. At the free-entry width of
Section~\ref{sec:numerics} for $g = 5$\,bp, a provider free to re-center inside
the range would do so a third of the half-width above the center, which lowers the
cost per unit of time by $4.4\%$, and at the lower boundary waiting is better
than re-centering.
The stationary equilibrium below is one of the restricted game.

\begin{proposition}[Profit rate and break-even]\label{prop:BEC}
The long-run profit rate of a provider of half-width $w$ against a stationary
aggregate $\Lbar$ is
\begin{equation}\label{eq:Pi}
  \Pi(w) := \Rcyc(w) - R_0(w),
  \qquad
  R_0(w) := \leff\Bigl[\tfrac12\varphi(w) + g\Bigr]
          = \frac{\sD^2\bigl(\varphi(w) + 2g\bigr)}{2w^2},
\end{equation}
so the width $w$ earns zero excess profit iff $\Rcyc(w) = R_0(w)$.
\end{proposition}
\begin{proof}
See Appendix~\ref{app:proofs}.
\qed
\end{proof}

The cost rate $R_0$ is the expected cost of a cycle, the boundary IL
$\tfrac12\varphi(w)$ (either boundary with probability $\tfrac12$, by the driftless
symmetry) plus the transaction cost $g$, spread over the expected cycle length
$1/\leff$. It is the rate that a position of half-width $w$ must earn to be
viable.

\begin{proposition}[Best-response width]\label{prop:wstar}
Let $\Rcyc(w) = B_\alpha/w^\alpha$ with $B_\alpha > 0$ and $\alpha \in (0,1]$ given
(Lemma~\ref{lem:RB} for $\alpha = 1$, Remark~\ref{rem:alpha} for $\alpha < 1$),
and treat $w$ as continuous on $(0,\infty)$. The profit
$\Pi(w) = B_\alpha w^{-\alpha} - R_0(w)$ has an interior maximizer $w^*$, which
satisfies
\begin{equation}\label{eq:FOC_exact}
  \frac{\alpha B_\alpha}{w^{\alpha+1}}
  = -\frac{\sD^2}{2} \frac{d}{dw}\!\left[\frac{\varphi(w) + 2g}{w^2}\right].
\end{equation}
For $w^*h \ll 1$ the critical point is unique and $\Pi''(w^*) < 0$.
\end{proposition}
\begin{proof}
See Appendix~\ref{app:proofs}.
\qed
\end{proof}

Proposition~\ref{prop:wstar} describes the best response to a given competition
level $B_\alpha$. With a coarse set of widths the best admissible width is one of
the two points of $\mathcal{W}$ adjacent to~$w^*$.

\begin{corollary}[leading-order value of $w^*$]\label{cor:wstar_asymp}
Let $\kz := h/4$. For small $w^*h$, so that $\varphi(w) \approx 2\kz w$, one
has
\[
  \Pi(w) \approx B_\alpha/w^\alpha - \kz\sD^2/w - \sD^2 g/w^2,
\]
and the first-order condition yields the closed forms below.
\begin{enumerate}
\item For $\alpha \in (0, 1)$ in the limit $g \to 0^+$,
\begin{equation}\label{eq:wstar_a}
  w^*_0 = \left(\frac{\kz\sD^2}{\alpha B_\alpha}\right)^{\!1/(1-\alpha)}.
\end{equation}
\item For $\alpha = 1$, $g > 0$ and $B_1 > \kz \sD^2$,
\begin{equation}\label{eq:wstar_c}
  w^*_0 = \frac{2g \sD^2}{B_1 - \kz\sD^2}
        = \frac{8 g \sigma^2}{h\,(\sigma_c^2 - \sigma^2)},
  \qquad \sigma_c := 2\sqrt{B_1 h}.
\end{equation}
\end{enumerate}
The exact $w^*$ differs from $w^*_0$ by a relative
$O\bigl((w^*_0h)^2 s^2/(1-s^2)\bigr)$, with $s = \sigma/\sigma_c$.
\end{corollary}
\begin{proof}
See Appendix~\ref{app:proofs}.
\qed
\end{proof}

The critical volatility $\sigma_c$ is the volatility at which fee income stops
covering the impermanent loss at every width. In terms of the \emph{participation
margin} $s := \sigma/\sigma_c$, \eqref{eq:wstar_c} reads
\begin{equation}\label{eq:nondim}
  w^*_0\,h = \frac{8\,g\,s^2}{1 - s^2}.
\end{equation}
The optimal range is narrow when fees dominate ($s \ll 1$) and widens as the pool
approaches its margin, since longer cycles spread the cost $g$ over more time. As
$\sigma \uparrow \sigma_c$ the leading-order width diverges, and for
$\sigma \ge \sigma_c$ no width is profitable at this $B_1$.

Free entry into the pool from the state $w = 0$ asks for a width at which profit
is both maximal and zero. At leading order no such width exists. For $\alpha = 1$
the best response~\eqref{eq:wstar_c} earns
\[
  \Pi(w^*_0) = \frac{(B_1 - \kz\sD^2)^2}{4g\,\sD^2} > 0,
\]
so the two conditions cannot hold together. Free entry instead lowers $B_1$,
through the aggregate $\Lbar$ in~\eqref{eq:B1}, until the \emph{maximal} profit is
zero, at a margin where the next order in $wh$ must be kept. Since $\Lbar \mapsto s$ is continuous and strictly increasing
by~\eqref{eq:B1}, the population settles in one of three cases. If the background
liquidity $L_0$ alone gives a margin above the zero-profit margin $s_\star$ of
Proposition~\ref{prop:margin}, then no provider enters. If the margin with every provider active is below
$s_\star$, then all providers are active and earn a positive profit. Otherwise a
fraction of providers is active, and the aggregate adjusts until $s = s_\star$ and
every provider, active or out, earns zero excess profit. The last case is the one
that Section~\ref{sec:numerics} tests.

\begin{proposition}[Free-entry margin]\label{prop:margin}
Let $\alpha = 1$, with $\Rcyc$ and $R_0$ in their exact forms
$\Rcyc = f\Vbar/(\Lbar\,\varepsilon_w)$ and~\eqref{eq:Pi}. The maximal profit
$\max_w \Pi(w)$ changes sign at a single margin $s_\star$, positive below it and
negative above it, with
\begin{equation}\label{eq:sstar}
  s_\star = \Bigl(1 + \tfrac74\, x^{*2}\Bigr)^{-1/2} + O(g),
  \qquad x^* := \frac{w^*h}{2} = \Bigl(\frac{12g}{7}\Bigr)^{1/3},
\end{equation}
and $w^* = (2/h)(12g/7)^{1/3}$ is the width held at the margin $s_\star$.
\end{proposition}
\begin{proof}
See Appendix~\ref{app:proofs}.
\qed
\end{proof}

The constants $\tfrac74$ and $\tfrac{12}{7}$ are exact. Solving the two conditions
$\Pi = 0$ and $\Pi' = 0$ numerically with the exact functions shows that the
neglected terms raise $s_\star$ by about $0.45\,g$, less than $5\times10^{-4}$
for $g \le 10$\,bp. The margin therefore sits strictly below one, by
$\tfrac78(12g/7)^{2/3}$ at leading order, an amount that vanishes with the cost.
At $g = 5$\,bp, $x^* \approx 0.095$ and $s_\star \approx 0.992$.

\begin{corollary}[entry rule and exclusion]\label{cor:entry}
Under Assumption~\ref{ass:boundary} and the decomposition~\eqref{eq:gsplit}, and at
the order of Proposition~\ref{prop:margin}, a provider of capital $c$ can profitably
supply a pool of margin $s$ at some width iff $g_0/c + \kappa < g_{\max}(s)$, where
\begin{equation}\label{eq:gmax}
  g_{\max}(s) = \frac{7}{12}\left[\frac{2}{\sqrt7}\sqrt{s^{-2}-1}\,\right]^{3}
  \quad (s < 1),
\end{equation}
so the minimum viable capital is $c_{\min}(s) = g_0/(g_{\max}(s) - \kappa)$. No
capital suffices once
\begin{equation}\label{eq:sexcl}
  s \;\ge\; s_{\mathrm{excl}} := s_\star(\kappa)
  = \Bigl(1 + \tfrac74 \bigl(\tfrac{12\kappa}{7}\bigr)^{2/3}\Bigr)^{-1/2} < 1 .
\end{equation}
\end{corollary}
\begin{proof}
The map $g \mapsto s_\star(g)$ of~\eqref{eq:sstar} is continuous and strictly
decreasing, and solving $s = s_\star(g)$ for $g$ gives~\eqref{eq:gmax}. By
Proposition~\ref{prop:margin}, provision pays at some width iff
$s < s_\star(g)$, that is iff $g < g_{\max}(s)$. Substituting
$g = g_0/c + \kappa$ and solving for $c$ gives $c_{\min}$, which is positive and
finite iff $g_{\max}(s) > \kappa$, that is iff $s < s_\star(\kappa)$.
\qed
\end{proof}

Because the proportional part $\kappa$ cannot be outgrown, past $s_{\mathrm{excl}}$
no boundary-triggered provider should supply the pool, however large. A pool whose
measured margin exceeds one is therefore excluded whatever the value of $\kappa$,
since $s_{\mathrm{excl}} < 1$ for every $\kappa > 0$.
Section~\ref{subsec:exclusion} evaluates this on data.

\subsection{Explicit value function}\label{sec:Vstar}

\begin{theorem}[Explicit $V^*$ in the fast-rebalancing regime]\label{thm:Vstar}
Under Assumptions~\ref{ass:Gamma} and~\ref{ass:boundary}, with half-width $w^*$,
aggregate $\Lbar$ and locally flat volume, the relative value function of the
exact dynamics inside the range ($|n| \le w^*$) is the downward parabola
\begin{equation}\label{eq:Vstar}
  \boxed{V^*(n)
    = -\frac{\varphi(w^*) + 2g}{2 (w^*)^2}\,n^2
    - \frac{\eta(w^*)}{2 w^*}\,n + C,}
\end{equation}
with $\varphi, \eta$ from~\eqref{eq:phi_eta}, a free constant $C$, and long-run gain
equal to the profit rate $\Pi(w^*)$, which is zero at the free-entry margin.
\end{theorem}
\begin{proof}
See Appendix~\ref{app:proofs}.
\qed
\end{proof}

The stationary equation fixes $V^*$ up to the constant $C$, and with the value of a
provider out of the pool set to zero every $C \in [0,g]$ is consistent with the
equilibrium. No continuum approximation enters, since the
discrete second difference of $n^2$ is exactly $2$ and that of $n$ is $0$. Outside the range the provider rebalances at once. The parabola has its vertex
to the left of $n = 0$ because of the asymmetry $\eta > 0$. The lower boundary is
the cheaper one, so the most valuable offset is slightly below the center.

\subsection{Stationary distributions and the equilibrium}\label{sec:mustar}\label{subsec:equilibrium}

Two stationary laws describe one position over time, one for its offset and one
for its center. Neither is a cross-section of the population at a date, since the
offsets of all providers move with the same price.

\begin{proposition}[Stationary offset law]\label{prop:offset}
Under Assumptions~\ref{ass:Gamma} and~\ref{ass:boundary} with half-width $w^*$, the
unique stationary law of the offset is the triangular law $\rho_{w^*}$
of~\eqref{eq:rho}, and positions are re-centered at the rate $\leff = \sD^2/(w^*)^2$.
\end{proposition}
\begin{proof}
See Appendix~\ref{app:proofs}.
\qed
\end{proof}

The offset law relaxes on the time scale of one cycle. The slowest mode of the
reset walk is odd in $n$, so the re-injection at the origin does not affect it, and
the spectral gap is
\[
  \sD^2\Bigl(1 - \cos\frac{\pi}{w^*}\Bigr)
  = \frac{\pi^2}{2}\,\leff\,\bigl(1 + O\bigl((w^*)^{-2}\bigr)\bigr).
\]
All economic quantities of Sections~\ref{subsec:stationary}--\ref{sec:Vstar}
depend on the offset alone and are therefore governed by this fast rate. The law of
the center of a tracked position mixes much more slowly, at the rate
$O(\sD^2/M^2)$ of the price walk on the whole grid, and it is uniform in the long
run up to boundary effects (Section~\ref{subsec:layers}). The results of this
section combine into the stationary equilibrium.

\begin{theorem}[Stationary equilibrium]\label{thm:MFE_triple}
Under Assumptions~\ref{ass:Gamma} and~\ref{ass:boundary}, let $\Lbar$ put the pool at
the margin $s_\star$ of Proposition~\ref{prop:margin} and let $w^*$ maximize $\Pi$.
Then the value $V^*$ of~\eqref{eq:Vstar} with $C \in [0,g]$, the offset law $\rho_{w^*}$
and the boundary rule $u^*$ at width $w^*$ form a stationary equilibrium within
boundary-triggered provision.
\end{theorem}

\begin{proof}
See Appendix~\ref{app:proofs}.
\qed
\end{proof}

\section{Numerical illustration: the participation margin on-chain}\label{sec:numerics}

This section takes the closed forms of Section~\ref{subsec:stationary} to data.
The object under test is the participation margin $s$.

\subsection{From the model to observables}\label{subsec:observables}

By~\eqref{eq:B1}, the critical volatility of Corollary~\ref{cor:wstar_asymp} is
\begin{equation}\label{eq:sigmac_obs}
  \sigma_c^2 = 4hB_1 = \frac{4f\Vbar}{\sqrt{\bar q}\,\Lbar},
  \qquad s = \frac{\sigma}{\sigma_c},
\end{equation}
with $\bar q$ the price level and $\Lbar$ the liquidity at the active tick. Every
input is posted or observable on-chain, nothing is fitted, and free entry predicts
$s = s_\star$, slightly below one.

We take $\sigma$ to be the realized volatility of hourly log closes. Because hourly
volume and liquidity move together, $B_1$ must be estimated pathwise, as the mean
of the hourly ratio $f V_t/(\sqrt{q_t}\,h\,L_t)$. Replacing it by a ratio of
separate window aggregates biases $B_1$ by up to a factor of two.

The model holds $\sigma$, $\Vbar$ and $\Lbar$ constant, and in the data they are
treated as constant within each $30$-day window. At the free-entry widths a cycle
lasts $25$ to $70$ days (Section~\ref{subsec:stationary}), so $s$ measured in one
window compares two contemporaneous rates rather than the outcome of completed
rebalancing cycles.

\subsection{Data}\label{subsec:data}

We use hourly pool state and the complete Mint, Burn and Collect event history of
six WETH/USDC pools on two chains and four protocol implementations, from January
2024 to July 2026 over each pool's lifetime. On Ethereum these are the Uniswap~V3
pools at the $0.3\%$ and $0.05\%$ tiers, and on Base the Uniswap~V3 ($0.05\%$),
Aerodrome ($0.0452\%$), PancakeSwap~V3 ($0.01\%$) and SushiSwap~V3
($0.01\%$) pools. The margin is reported for five of the six. The SushiSwap pool
holds two orders of magnitude less liquidity than the others, so a single position
is a non-negligible share of its aggregate and the mean-field reading does not
apply.

Hourly pool state comes from the subgraph, with $22340$ and $22524$
observations on the two Ethereum pools. Swap-level data come from an independent
index of the chain and exist for Base only, $22$ million swaps over $29988$
pool-hours. Position events cover $25161$ pool-hours on Ethereum and $26001$
on Base. Two internal checks fix the chain of units on which every number depends.
The effective fee rate, fees divided by volume, is $0.3\%$ and $0.05\%$ on the
two Ethereum pools, equal to the posted tiers to six decimal places. And $\Vbar$ measured from the volume
series agrees to within $2\%$ with the same object recovered from per-liquidity
fee-growth increments, an independent path through the protocol's own accounting.

\subsection{The margin holds, and it moves with volatility}\label{subsec:margin}

\begin{figure}[t]
\centering
\includegraphics[width=0.8\textwidth]{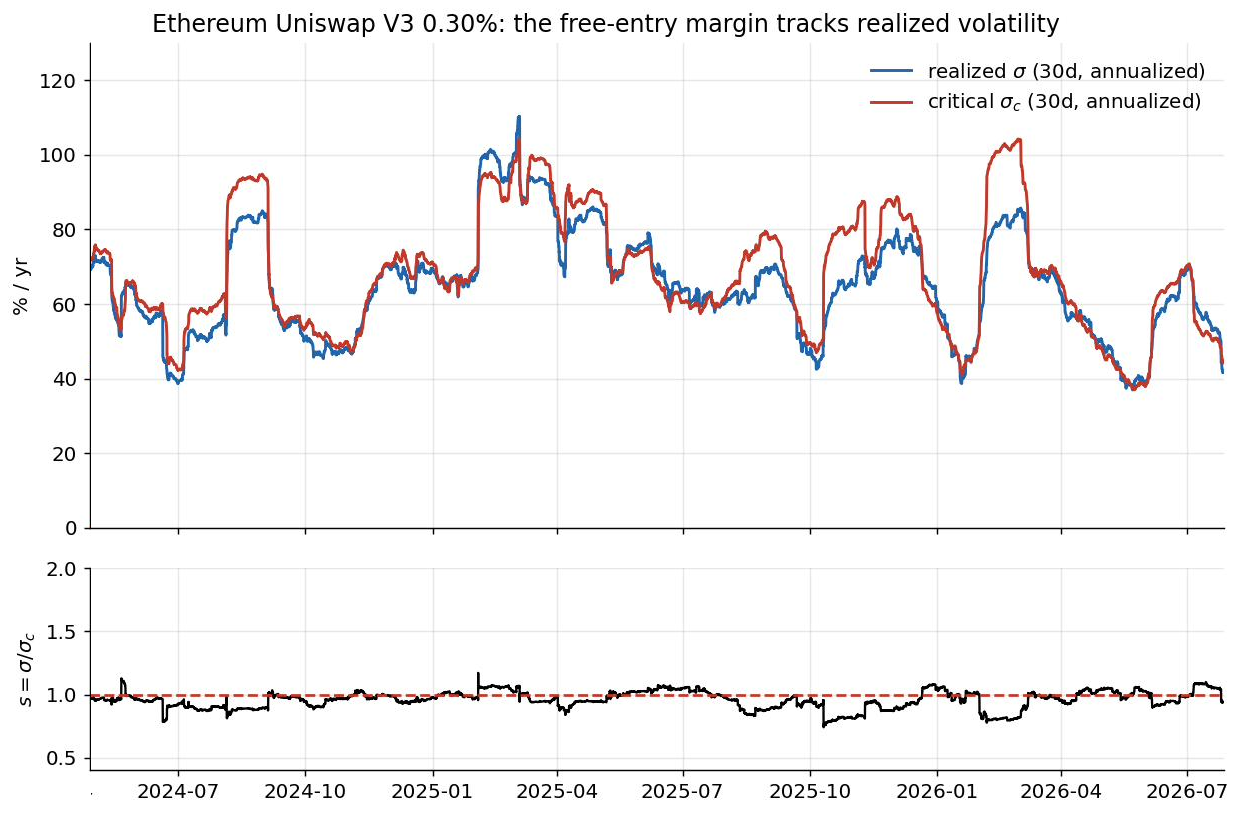}
\caption{Ethereum Uniswap~V3 $0.3\%$. The top panel shows the $30$-day annualized
realized $\sigma$ and the critical $\sigma_c$ of~\eqref{eq:sigmac_obs}, neither
fitted to the other, and the bottom panel the margin $s = \sigma/\sigma_c$. Both series begin once the
$30$-day window is full and the pool's liquidity history covers it.}\label{fig:E1}
\end{figure}

Figure~\ref{fig:E1} is the strongest form of the result. On the Ethereum $0.3\%$
pool, which pays neither token emissions nor a protocol fee, the $30$-day
$\sigma_c$ tracks the $30$-day realized $\sigma$ through a twofold swing, from
$41\%$ to $85\%$ annualized, for ten consecutive quarters, with $s$ inside roughly
$0.9$--$1.1$ throughout. We compute $95\%$ percentile confidence intervals for $s$
from $2000$ replicates of a circular block bootstrap with one-week blocks ($168$ hourly
observations). Each replicate resamples the inputs and rebuilds $s$, because $s$ is
a ratio of two means. This gives $s = 0.983$ for the $0.3\%$ pool, with $95\%$ confidence interval
$[0.945,\,1.023]$, and $s = 1.059$ for the $0.05\%$ pool, with $95\%$ confidence
interval $[1.016,\,1.103]$. The intervals are stable to within $0.01$ with daily
or monthly blocks.
For every position size above $\$30$k the free-entry margin~\eqref{eq:sstar} of the
$0.3\%$ pool lies between $0.978$ and $s_{\mathrm{excl}} = 0.982$
(Section~\ref{subsec:exclusion}), inside the first interval. The second interval
excludes both one and every free-entry margin of the $0.05\%$ pool, which lie below
$0.993$, so that pool sits above its margin by more than sampling error.

\begin{figure}[t]
\centering
\includegraphics[width=0.8\textwidth]{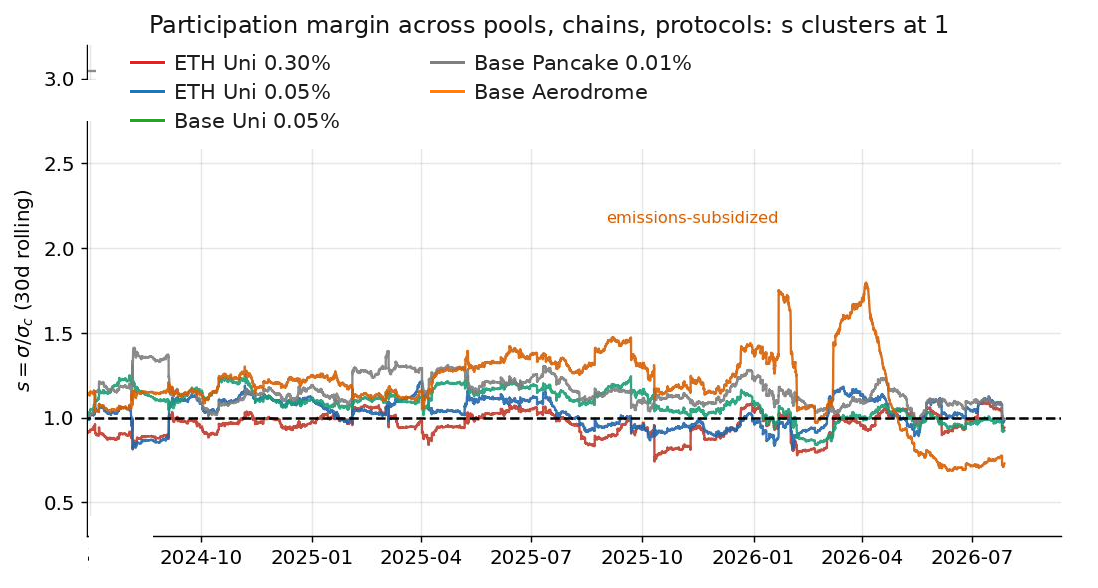}
\caption{Thirty-day rolling $s$ across five pools, two chains and four protocols,
each from the end of its warm-up window. The emissions-subsidized pool
floats above, most visibly in early 2026 when emissions rose. For the
protocol-fee pool the fee-only estimate understates the margin that providers
face.}\label{fig:E2}
\end{figure}

Across pools (Figure~\ref{fig:E2}) the series cluster at one. Over the full
samples the margin is $0.98$ on the Ethereum $0.3\%$ pool, $1.05$ on the Base
Uniswap pool, $1.06$ on the Ethereum $0.05\%$ pool, $1.15$ on Aerodrome and $1.18$
on PancakeSwap. The two largest deviations have the sign the cost channel predicts. Aerodrome pays
gauge emissions on top of fees, a subsidy that sustains provision above the
fee-only margin. PancakeSwap diverts part of the swap fee to the protocol before it
reaches providers. The swap-level data record the share reaching providers
directly, $67\%$ of the fee, identical on the two token sides to four decimal
places, and applying it moves the LP-effective margin to $1.44$. The two Uniswap $0.05\%$ pools
carry neither emissions nor a protocol fee and still sit above one, which the cost
channel alone does not explain. They are close to each other, $1.05$ and $1.06$, on
chains whose gas costs differ by orders of magnitude and which share the tier, and
hence the proportional cost $\kappa$.

\subsection{The margin is a ratio of realized to implied volatility}\label{subsec:iv}

The value of liquidity $\Lbar$ over one tick band of log-width $h$ is
$\mathrm{TVL}_{\mathrm{tick}} \approx \sqrt{\bar q}\,h\,\Lbar/2$,
by~\eqref{eq:eps_centered} with $2wh = h$. The on-chain implied volatility of
\cite{lambert2021}, obtained by equating fee income to the time value of the
range, is $\sigma_{\mathrm{IV}} = 2f\sqrt{\Vbar/\mathrm{TVL}_{\mathrm{tick}}}$.
By~\eqref{eq:sigmac_obs},
\begin{equation}\label{eq:iv}
  \sigma_c = \sqrt{\frac{2f\Vbar h}{\mathrm{TVL}_{\mathrm{tick}}}}
  = \sqrt{\frac{h}{2f}}\;\sigma_{\mathrm{IV}},
\end{equation}
so $\sigma_c = \sigma_{\mathrm{IV}}$ on the Uniswap volatile tiers, where $h = 2f$
up to a relative $5\times10^{-5}$ (for instance $h = 60\ln 1.0001$ against
$2f = 0.006$, Section~\ref{sec:intro}). Both quantities are the volatility at which the
fee income of the active liquidity equals its loss-versus-rebalancing
(Section~\ref{subsec:leff}), the critical volatility on the actual grid and the
formula of \cite{lambert2021} with $h = 2f$ built in. The relation is thus a
property of a protocol's grid, and the sample shows both sides of that. On the
$0.01\%$ tier $h \approx f$ and $\sigma_c = \sigma_{\mathrm{IV}}/\sqrt2$, and Aerodrome
pairs a $4.52$\,bp fee with $h = 100\ln 1.0001$, giving $h/f = 22$ and
$\sigma_c \approx 3.3\,\sigma_{\mathrm{IV}}$.

The margin is thus a ratio of realized to implied volatility, and $s \approx 1$
says that concentrated-liquidity fees price realized variance correctly on
average, under a risk-neutral margin. Whether the residual is a vanishing variance
risk premium or a premium offset by income the model does not price, these data do
not settle.

A mechanical account reproduces $s \approx 1$ with no provider entering or
leaving. If the whole volume is arbitrage that realigns the pool whenever its price
leaves a band of half-width $f$ in logarithms around the reference price, then in
the limit of fast blocks the fees paid tend to the loss-versus-rebalancing
$\sigma^2\sqrt{\bar q}\,\Lbar/4$ of the active liquidity \cite{mmrfees}, which
by~\eqref{eq:sigmac_obs} is $s = 1$ at every depth and on every tier, since
volume is then proportional to liquidity. The level of the margin therefore does
not separate this account from free entry, and a test needs the response of
liquidity to changes in volume. The comparison across tiers rejects only the
coarser account in which arbitrage trades once per tick crossed, which gives
$s^2 \propto h/f$ and puts the $0.01\%$ tier lowest of the five pools, while it
is measured highest. The measured liquidity is that of persistent positions,
since positions minted and burned in the same block, which include just-in-time
liquidity, make up $23\%$ and $43\%$ of mint events on the two Ethereum pools but
only $0.05\%$ and $1.5\%$ of the liquidity minted, and below $1\%$ of the
liquidity on three of the four Base pools.

\subsection{A hard entry cutoff}\label{subsec:exclusion}

Corollary~\ref{cor:entry} turns the margin into a decision rule with a hard
cutoff. The proportional cost $\kappa$ of~\eqref{eq:gsplit} is the cost of restoring the
range after an exit. At that moment the position is held in one token, and
re-centering converts about half of it, so $\kappa$ is half the cost of one swap,
the fee $f_{\mathrm{ex}}$ of the venue used for the conversion plus a slippage
allowance $\varsigma$. We set $f_{\mathrm{ex}}$ to the pool's own fee tier, which
assumes that the provider converts in the pool it supplies, and $\varsigma = 5$\,bp. Both are held constant over the sample, and
neither is estimated from the data. On Base the measured median swapped share is
$0.28$ rather than one half ($0.31$ across pools), and $44\%$ of the matched swaps
route through other pools, so $\kappa$ as set here is an upper bound. With the
measured share the Ethereum $0.3\%$ pool has $\kappa = 9.8$\,bp and
$s_{\mathrm{excl}} = 0.988$, and at $s = 0.983$ it is open from about \$30k. The fixed part is $g_0 = \$20$ on Ethereum and
a few cents on Base, also held constant, and the minimum capital below scales
linearly with it.
Table~\ref{tab:exclusion} evaluates $s_{\mathrm{excl}}$ for each pool.

\begin{table}[!ht]
\centering
\caption{Exclusion margins. The proportional cost $\kappa$ is that of~\eqref{eq:gsplit},
$s_{\mathrm{excl}}$ the margin~\eqref{eq:sexcl} past which no capital suffices, and
$s$ the full-sample margin of Section~\ref{subsec:margin}.}
\label{tab:exclusion}
\small
\setlength{\tabcolsep}{5.5pt}
\begin{tabular}{@{}lrrrl@{}}
\toprule
pool & $\kappa$ & $s_{\mathrm{excl}}$ & $s$ observed & verdict \\
\midrule
Ethereum $0.3\%$         & $17.5$\,bp & $0.9823$ & $0.983$ & at the margin \\
Base Uniswap $0.05\%$     & $5$\,bp    & $0.9922$ & $1.05$ & closed at every capital \\
Ethereum $0.05\%$         & $5$\,bp    & $0.9922$ & $1.06$ & closed at every capital \\
Aerodrome                 & $4.76$\,bp & $0.9924$ & $1.15$ & closed at every capital \\
PancakeSwap, LP-effective & $3$\,bp    & $0.9944$ & $1.44$ & closed at every capital \\
\bottomrule
\end{tabular}
\end{table}

Since $s_{\mathrm{excl}} < 1$ for every $\kappa > 0$, a pool measured above one
is closed whatever the proportional cost, so only the Ethereum $0.3\%$ verdict
uses the cost estimate. That pool sits at its exclusion margin. Its point estimate
$s = 0.983$ lies just above the leading-order $s_{\mathrm{excl}} = 0.9823$ and just
below the exact one, $0.9831$, at which the pool is open only above a capital of
about \$2.2M, and its confidence interval $[0.945,\,1.023]$ covers both. Inside the
interval the minimum viable capital $c_{\min}$ of
Corollary~\ref{cor:entry} is about $\$56$k at $s = 0.98$, rises to $\$130$k at
$s = 0.9813$ and diverges at $s_{\mathrm{excl}}$. These figures use the
leading-order margin~\eqref{eq:sstar}. With the exact margin in its place the same two
figures are about $\$38$k and $\$68$k, so the minimum capital is an order of
magnitude rather than a target. It is also sensitive to $\kappa$.
Routing the conversion through the $0.05\%$ pool of the same pair lowers $\kappa$
from $17.5$ to $5$\,bp, raises $s_{\mathrm{excl}}$ from $0.982$ to $0.992$, and
opens the pool at the point estimate $s = 0.983$ from a capital of about \$18k at
leading order.
All five verdicts hold within boundary-triggered
provision.

A backtest on the four Base pools agrees by another route. Over the $7497$ hours from
September 2024 to July 2025 covered by the swap-level data, a strategy that provides
only when the trailing margin permits is in the market $2.5$--$12\%$ of hours and still does not
turn a profit. None of this says that provision is irrational. Providers earning
emissions, internalizing their own order flow, or holding inventory for unrelated
reasons face a different $g$ than the one priced here.

\subsection{What the diffusion misses}\label{subsec:diffusion}

At the costs above, the free-entry half-width is $w^*h = 2x^* \approx 0.19$ on
Base ($g = 5$\,bp) and $0.31$ on the Ethereum $0.3\%$ pool at $c = \$56$k
($g = 21$\,bp). At the $72.5\%$ annualized volatility measured on Base, the
loss-versus-rebalancing of a full-range position is $\sigma^2/8 \approx 6.6\%$ of
capital per year, and by~\eqref{eq:lvr} the rate at these widths is $40$ to $70\%$
per year. Fee income has to cover that, so the break-even condition is a real
constraint.

The diffusive benchmark does well on the frequency of range exits and less well on
their cost. The four Base pools share one price, so every estimator of its
volatility must agree across them.
Close-to-close estimates do, at $72\%$ to $74\%$ annualized at hourly,
four-hourly and daily sampling on every pool, while swap-level estimates spread
over twenty percentage points.
On the hourly closes, bipower variation attributes about $15\%$ of total variance to
jumps, the same on all four pools to within a percentage point. Replaying the
reset policy against their realized tick path, the observed rate at which a position
leaves its range matches the diffusive first-passage rate $\leff = \sigma^2/(wh)^2$
to within $1\%$ at half-widths of $8\%$ and above, and to within $10\%$ at $4\%$.
Yet always-on provision loses $6$--$18\%$ of capital per year on the Ethereum
pools, and on the $0.3\%$ pool, which sits at its margin, a diffusive model prices
it near zero. If the frequency of exits is right there as it is on Base, the gap
must come mostly from the cost of each exit, an overshoot cost incurred because a
jump carries the price past the boundary rather than to it. It is the term a
jump--diffusion treatment would have to supply.

\section{Conclusion}\label{sec:conclusion}

We have built a mean-field model of liquidity provision on concentrated-liquidity
exchanges that is solvable in closed form at its stationary equilibrium. In an
averaged version of the game an equilibrium exists and its payoff-relevant part is
unique, for identical providers and for finitely many cost types alike. A critical volatility built from
observable pool quantities separates viable from unviable provision, free entry
holds a pool at a margin just below it, and on five WETH/USDC pools the critical
volatility tracks realized volatility, with the measured margin of the Ethereum
$0.3\%$ pool at its free-entry value within sampling error.

The results rest on simplifications that should be kept in view. The price is
exogenous and follows a GBM, the order flow is exogenous and stationary, and
providers are risk neutral and differ at most in their cost. The model therefore
rules out price impact of liquidity provision and endogenous arbitrage frictions,
and it describes an idealized venue rather than on-chain execution in detail. The
general theorems concern the averaged game, and the closed forms concern
boundary-triggered provision. The constants of Proposition~\ref{prop:margin}
belong to the driftless log-price. With a log-drift of $-\theta\sigma^2/2$,
$0 \le \theta \le 1$, the $7$ in~\eqref{eq:sstar} becomes
$7 - 12\theta + 4\theta^2$, which is negative for $\theta > 0.79$, and for a
martingale price ($\theta = 1$) wide ranges are profitable at $s = 1$ and the
bound $s_{\mathrm{excl}} < 1$ is lost. Open directions include jump--diffusion
and stochastic-volatility prices, which Section~\ref{subsec:diffusion} shows would
change the cost of a boundary crossing more than its frequency, endogenous and
time-varying order flow and price impact, a Stackelberg problem of fee design by the protocol, and
the unrestricted rebalancing problem, including asymmetric ranges.

%
%

\appendix
\makeatletter
\renewcommand{\theHsection}{appendix.\Alph{section}}
\makeatother
\section{Proofs of the main results}\label{app:proofs}

\paragraph{Exact impermanent loss \cite{tang}.}
Write $\mathrm{clip}_{[a,b]}(q) := \max\{a,\min\{b, q\}\}$ for the price clamped to the range
and $\varepsilon_{a,b}(q)$ for the per-unit cost of liquidity
(Definition~\ref{def:pos}). The exact IL of a position $(j,w)$ when the price moves
from $q$ to $q'$ is, as a fraction of unit capital,
\begin{equation}\label{eq:ILexact}
  \ILe(j, w, q', q) := \frac{\hat\Theta_{a,b}(q', q)}{\varepsilon_{a,b}(q)},
  \qquad a = t_{j-w},\quad b = t_{j+w},
\end{equation}
where, with $\bar q := \mathrm{clip}_{[a,b]}(q)$ and $\bar q' := \mathrm{clip}_{[a,b]}(q')$,
\begin{equation}\label{eq:tauhat}
  \hat\Theta_{a,b}(q', q) := \sqrt{\bar q} + \frac{q'}{\sqrt{\bar q}}
  - \sqrt{\bar q'} - \frac{q'}{\sqrt{\bar q'}} .
\end{equation}
It is nonnegative \cite{tang}. For a \emph{centered} position, with $q = t_j$ and
an in-range price $q' = t_je^{nh}$, it reduces to
\begin{equation}\label{eq:ILexact_inrange}
  \ILe(w,n) := \ILe(j, w, t_je^{nh}, t_j)
  = \frac{\bigl(1 - e^{nh/2}\bigr)^2}{2\bigl(1 - e^{-wh/2}\bigr)},
  \qquad |n| \le w .
\end{equation}
The reduction uses $\sqrt{a} + t_j/\sqrt{b} = 2\sqrt{t_j}\,e^{-wh/2}$, which holds
at the center only. Evaluating~\eqref{eq:ILexact_inrange} at $n = \pm w$ gives
the boundary forms~\eqref{eq:ILrebal} and their sum and difference~\eqref{eq:phi_eta}.

\begin{proof}[Proof of Lemma~\ref{lem:occupation}]
The expected occupation times $G_w(n)$ of the walk with generator
$\mathcal{L} = \tfrac12\sD^2\Delta_d$, started at $0$ and absorbed at $\pm w$, solve
the discrete Poisson problem $\mathcal{L} G_w = -\delta_0$ on $|n| < w$ with
$G_w(\pm w) = 0$, since $\mathcal{L}$ is symmetric. The piecewise-linear candidate
$G_w(n) = (w - |n|)/\sD^2$ satisfies $\Delta_d G_w(n) = 0$ for $0 < |n| < w$ and
$\tfrac12\sD^2\bigl[G_w(1) + G_w(-1) - 2G_w(0)\bigr] = \tfrac12\sD^2 \cdot
(-2/\sD^2) = -1$ at $n = 0$, with the boundary values $0$. By uniqueness of the
solution of the discrete Dirichlet problem it is the Green's function. Summing,
$\sum_{n=-w}^{w} (w - |n|) = (2w+1)w - w(w+1) = w^2$, so
$\sum_n G_w(n) = w^2/\sD^2 = \E[\tau_{\mathrm{exit}}]$, and
$\rho_w := \leff G_w$ is a probability vector.
\qed
\end{proof}

\begin{proof}[Proof of Lemma~\ref{lem:simplex}]
The rows of every $Q^\nu$ sum to zero, so summing~\eqref{eq:KFE} over $(k,\tilde w)$
gives $\sum\mu_t \equiv 1$. The equation is linear, $\tfrac{d}{dt}\mu_t = \mu_t q_t$, with
rate matrices $q_t$ whose off-diagonal entries are nonnegative and whose diagonal
entries are bounded by $\Gamma$. Then $\mu_t e^{\Gamma t}$ solves a linear equation
with the entrywise nonnegative matrix $q_t + \Gamma I$, and the Picard iteration
from $\mu_0 \ge 0$ preserves nonnegativity, so $\mu_t \ge 0$.
\qed
\end{proof}

\begin{proof}[Proof of Theorem~\ref{thm:exist}]
Write $z = (j,w)$ for a state and, for a strict control $u$,
\[
  \mathcal{K}(z,V,u) := \sum_{z'} Q^u_{zz'}V(z') - \sum_{z' \ne z} Q^u_{zz'}\,c(z,z'),
\]
extended linearly to mixed controls, so that \eqref{eq:HJB} reads
\[
  -\frac{d}{dt}V(z) = \Rfee(z,\mu_t) + \mathcal{H}(z,V),
  \qquad \mathcal{H}(z,V) := \max_{u} \mathcal{K}(z,V,u).
\]

\textbf{(1) Invariant set.} The off-diagonal entries of a row of $Q^\nu$ sum to at
most $\Gamma$ and the diagonal entry is at most $\Gamma$ in absolute value, so every
$Q^\nu$ has absolute row sums at most $2\Gamma$. In
$\mathcal{X} := \mathcal{C}([0,T]; \R^{|\mathcal{S}|})$ the set $K_0$ of paths
$\mu$ with $\mu_t \in \Sigma_{|\mathcal{S}|}$ and
$\norm{\mu_t-\mu_s}_1 \le 2\Gamma|t-s|$ is nonempty, convex, closed and uniformly
bounded, hence compact by Arzel\`a--Ascoli~\cite{kolmogorovfomin}.

\textbf{(2) HJB.} Fix $\mu \in K_0$. Since $L(k,\mu) \ge L_0 > 0$, the map
$t \mapsto \Rfee(z,\mu_t)$ is continuous. The map $V \mapsto \mathcal{H}(z,V)$ is a
maximum of finitely many affine maps, each Lipschitz in the sup norm with constant
$2\Gamma$. The terminal-value problem for~\eqref{eq:HJB} therefore has a unique
$C^1$ solution $V^\mu$, and by Gronwall's inequality
\[
  \sup_t\|V^\mu_t - V^{\mu'}_t\|_\infty \le T e^{2\Gamma T}
  \sup_{t,z}|\Rfee(z,\mu_t) - \Rfee(z,\mu'_t)|,
\]
so $\mu \mapsto V^\mu$ is continuous.

\textbf{(3) Best responses and the KFE.} For each $(t,z)$ the set
$\mathcal{N}^*(t,z,\mu)$ of the $\nu \in \mathcal{P}(\mathcal{U})$ supported on
$\arg\max_u \mathcal{K}(z,V^\mu_t,u)$ is a nonempty face of the simplex
$\mathcal{P}(\mathcal{U})$, hence convex and compact. Choosing the first maximizer
in a fixed order of the finite set $\mathcal{U}$ gives a measurable selection,
since each $t \mapsto \mathcal{K}(z,V^\mu_t,u)$ is continuous. For every
measurable selection $\nu_t(z) \in \mathcal{N}^*(t,z,\mu)$, \eqref{eq:KFE} is a
linear ODE with bounded measurable coefficients, which by Carath\'eodory's
theorem~\cite{coddington1955} has a unique absolutely continuous solution from
$\mu_0$. The solution stays in the simplex by Lemma~\ref{lem:simplex} and satisfies
$\|\tfrac{d}{dt}\hat\mu_t\|_1 \le 2\Gamma$, so it lies in $K_0$. Let $\Phi(\mu)$ be the
set of these solutions over all measurable selections.

\textbf{(4) Convex values.} Let $\hat\mu^1, \hat\mu^2 \in \Phi(\mu)$ with controls
$\nu^1, \nu^2$, and let $\lambda \in [0,1]$. Because $Q^\nu$ is linear in $\nu$, the
path $\hat\mu = \lambda\hat\mu^1 + (1-\lambda)\hat\mu^2$ solves~\eqref{eq:KFE}
with the state-wise mixture
\[
  \nu_t(z) = \frac{\lambda\hat\mu^1_t(z)\,\nu^1_t(z)
  + (1-\lambda)\hat\mu^2_t(z)\,\nu^2_t(z)}{\hat\mu_t(z)},
\]
and with an arbitrary element of $\mathcal{N}^*$ where $\hat\mu_t(z) = 0$. By
convexity $\nu_t(z) \in \mathcal{N}^*(t,z,\mu)$, so $\hat\mu \in \Phi(\mu)$.

\textbf{(5) Closed graph.} Let $\mu^m \to \mu$ and $\hat\mu^m \to \hat\mu$ in $K_0$
with $\hat\mu^m \in \Phi(\mu^m)$ generated by $\nu^m$. The fluxes
$F^m_t(z,u) := \hat\mu^m_t(z)\,\nu^m_t(z)(u)$ are bounded and measurable, so along
a subsequence $F^m \to F$ weakly-$*$ in $L^\infty$. The limit satisfies
$\sum_u F_t(z,u) = \hat\mu_t(z)$, and $\hat\mu$ solves the linear equation driven
by $F$. Each $F^m$ is supported on maximizers, that is
\[
  \int_0^T\sum_{z,u} F^m_t(z,u)\,\bigl[\mathcal{H}(z,V^{\mu^m}_t)
  - \mathcal{K}(z,V^{\mu^m}_t,u)\bigr]\,dt = 0,
\]
and since $V^{\mu^m} \to V^\mu$ uniformly by (2), the identity holds in the limit.
Hence $F_t(z,\cdot)$ is supported on the maximizers for $V^\mu$, and
$\nu_t(z) := F_t(z,\cdot)/\hat\mu_t(z)$ is an admissible selection, so
$\hat\mu \in \Phi(\mu)$. In particular the values of $\Phi$ are closed subsets of
$K_0$, hence compact.

\textbf{(6) Fixed point.} By the Kakutani--Fan--Glicksberg theorem
\cite{aliprantis2006} in the locally convex space $\mathcal{X}$, there is
$\mu^* \in \Phi(\mu^*)$, and the triple $(V^{\mu^*}, \mu^*, \nu^*)$, with $\nu^*$
the selection generating $\mu^*$, is an MFE in the sense of
Definition~\ref{def:MFE}.
\qed
\end{proof}

\begin{proof}[Proof of Theorem~\ref{thm:unique}]
Use the notation of the previous proof and write $R(z,\mu) := \Rfee(z,\mu)$. Let
$(V^1, \mu^1, \nu^1)$ and $(V^2, \mu^2, \nu^2)$ be two MFEs with
$\mu^1_0 = \mu^2_0$, and set
\[
  \Xi(t) := \sum_z [V^1 - V^2](t,z)\,[\mu^1 - \mu^2](t,z).
\]
Then $\Xi(0) = 0$ by the equal initial laws, and $\Xi(T) = 0$ because the terminal
condition does not depend on the population. The KFE gives
\[
  \sum_z [V^1-V^2](z)\,\tfrac{d}{dt}\mu^i(z)
  = \sum_z \mu^i(z)\,\bigl[\mathcal{K}(z,V^1,\nu^i(z)) - \mathcal{K}(z,V^2,\nu^i(z))\bigr],
\]
in which the costs cancel. Differentiating $\Xi$ and substituting the two HJB
equations gives
\[
  \frac{d\Xi}{dt} = -\sum_z [R(z,\mu^1) - R(z,\mu^2)][\mu^1 - \mu^2](z)
  + \sum_z \mu^1(z)\,D^2(z) + \sum_z \mu^2(z)\,D^1(z),
\]
with $D^i(z) := \mathcal{H}(z,V^i) - \mathcal{K}(z,V^i,\nu^{3-i}(z)) \ge 0$,
because $\nu^i$ attains the maximum $\mathcal{H}(z,V^i)$ and $\nu^{3-i}$ need not.
Integrating over $[0,T]$,
\begin{align*}
  &\int_0^T \sum_z [R(z,\mu^1_t) - R(z,\mu^2_t)][\mu^1_t - \mu^2_t](z)\,dt\\
  &\qquad = \int_0^T \sum_z \bigl(\mu^1_t(z)D^2_t(z) + \mu^2_t(z)D^1_t(z)\bigr)\,dt
  \;\ge\; 0 .
\end{align*}
By~\eqref{eq:monotone} the integrand on the left is $\le 0$ at every $t$, so both
sides vanish. Hence $L(k,\mu^1_t) = L(k,\mu^2_t)$ for every $k$ and a.e.\ $t$,
since $\pi_k f\mathcal{V}(k) > 0$, and for every $t$ by continuity. The two HJB
equations then have the same data, and $V^1 = V^2$ by uniqueness of their
solution. Each $\nu^i$ is supported on the maximizers for $V^i = V^1$. If the
maximizer is unique, then $\nu^1_t(z) = \nu^2_t(z)$ for a.e.\ $t$ and every $z$, so
$\mu^1$ and $\mu^2$ solve the same linear equation from the same initial law and
coincide.
\qed
\end{proof}

\begin{proof}[Proof of Proposition~\ref{prop:BEC}]
Under boundary-triggered rebalancing the offset is a regenerative process, which
restarts at $n = 0$ at each rebalancing. By the renewal--reward theorem
\cite{asmussen}, the long-run income rate is the expected income per cycle divided
by the expected cycle length $1/\leff = w^2/\sD^2$. By Lemma~\ref{lem:occupation}
the expected income per cycle is
\[
  \E\!\int_0^{\tau_{\mathrm{exit}}} r\,dt
  = \sum_{n=-w}^{w} G_w(n)\, r(j+n)
  = \frac{1}{\leff} \sum_{n=-w}^{w} \rho_w(n)\, r(j+n)
  = \frac{\Rcyc(w)}{\leff},
\]
the in-range rate of Definition~\ref{def:Rcyc}. By the driftless symmetry of the
offset, exit is equally likely at either boundary, so the expected cost per cycle
is $\tfrac12\varphi(w) + g$, and the long-run profit rate is $\Rcyc(w) - R_0(w)$.
\qed
\end{proof}

\begin{proof}[Proof of Proposition~\ref{prop:wstar}]
The condition $\Pi'(w) = 0$ is~\eqref{eq:FOC_exact}. As $w \to 0^+$ the term $-\sD^2 g/w^2$ of $-R_0$ drives $\Pi \to -\infty$, since it
dominates $B_\alpha w^{-\alpha}$ for $\alpha \le 1$, and as $w \to \infty$ the
exponential growth of $\varphi$ does the same, so an interior maximum exists. For
small $wh$, $\varphi(w) \approx 2\kz w$ gives $R_0(w) \approx \sD^2(\kz/w + g/w^2)$,
hence
\[
  R_0'(w) \approx -\sD^2\!\left(\frac{\kz}{w^2} + \frac{2g}{w^3}\right),
  \qquad
  R_0''(w) \approx \sD^2\!\left(\frac{2\kz}{w^3} + \frac{6g}{w^4}\right).
\]
At a critical point, $\alpha B_\alpha w^{-\alpha-1} = |R_0'(w)|$, so
\begin{align*}
  \Pi''(w) &= \alpha(\alpha+1)B_\alpha w^{-\alpha-2} - R_0''(w)
  = \frac{(\alpha+1)\,|R_0'(w)|}{w} - R_0''(w)\\
  &\approx \sD^2\Bigl[\frac{(\alpha-1)\kz}{w^3} + \frac{(2\alpha-4)\,g}{w^4}\Bigr],
\end{align*}
which is negative for $\alpha \le 1$ and $g > 0$. For $wh \ll 1$ every critical
point is therefore a strict local maximum, and a $C^1$ function on an interval
whose critical points are all strict local maxima has at most one.
\qed
\end{proof}

\begin{proof}[Proof of Corollary~\ref{cor:wstar_asymp}]
As $wh \to 0$, $\varphi(w) \approx 2\kz w$ by~\eqref{eq:phi_exp}, so
\[
  \Pi'(w) \approx -\alpha B_\alpha/w^{\alpha+1} + \kz\sD^2/w^2 + 2\sD^2 g/w^3 .
\]
Setting $\Pi'(w)=0$ and multiplying by $w^{\alpha+1}$ gives $\alpha B_\alpha =
\kz\sD^2 w^{\alpha-1} + 2\sD^2 g\, w^{\alpha-2}$. For $\alpha \in (0,1)$ and
$g \to 0^+$ this yields~\eqref{eq:wstar_a}. For $\alpha=1$ and $g>0$ it gives
$w_0^* = 2\sD^2 g/(B_1 - \kz\sD^2)$, and substituting $\kz = h/4$,
$\sD^2 = \sigma^2/h^2$ and $\sigma_c^2 := 4h B_1$ gives~\eqref{eq:wstar_c}. As for
the error, in $\varphi(w) = 2\kz w + \tfrac{(wh)^2}{8} + \tfrac{(wh)^3}{12} +
O((wh)^4)$ the quadratic term contributes the constant $\sD^2 h^2/16 = \sigma^2/16$
to $R_0$ and drops out of the first-order condition. The first surviving correction
comes from the cubic term, which adds $-\sD^2h^3/24$ to $\Pi'$. Measured against
the difference $B_1 - \kz\sD^2$ that sets $w^*_0$, it moves the root by the relative
amount $(w_0^*h)^3/(48g) = (w_0^*h)^2 s^2/(6(1-s^2))$, which is small when $s$ is
bounded away from one. Near the margin Proposition~\ref{prop:margin} applies
instead.
\qed
\end{proof}

\begin{proof}[Proof of Proposition~\ref{prop:margin}]
Write $x := wh/2$ and $A := f\Vbar/(2\sqrt{\bar q}\,\Lbar) = hB_1/2$, so that
$\Rcyc = A/(1-e^{-x})$ exactly by~\eqref{eq:eps_centered}, $\leff = \sigma^2/(4x^2)$,
and $\sigma_c^2 = 4hB_1 = 8A$, that is $s^2 = \sigma^2/(8A)$. The profit rate is
\[
  \Pi(x) = \frac{A}{1 - e^{-x}}
  - \frac{\sigma^2}{4x^2}\Bigl[\tfrac12\bigl(e^{x}-1\bigr)\cosh x + g\Bigr].
\]
We expand with
\[
  \frac{1}{1-e^{-x}} = \frac1x + \frac12 + \frac{x}{12} + O(x^3),
  \qquad
  \tfrac12(e^x-1)\cosh x = \frac{x}{2} + \frac{x^2}{4} + \frac{x^3}{3} + O(x^4),
\]
and write $A = \sigma^2/8 + \omega$. Then
\[
  \Pi(x) = \omega\Bigl(\frac1x + \frac12\Bigr)
  - \frac{7\sigma^2}{96}\,x - \frac{\sigma^2 g}{4x^2}
  + O\bigl(\omega x + \sigma^2x^2\bigr).
\]
We look for the pair $(x, \omega)$ with $\Pi = 0$ and $\Pi' = 0$, with
$\omega = O(\sigma^2x^2)$ and $g = O(x^3)$. At the leading order $O(\sigma^2 x^2)$,
the two conditions multiplied by $x$ and by $x^2$ read
\[
  \omega - \frac{7\sigma^2}{96}x^2 - \frac{\sigma^2 g}{4x} = 0,
  \qquad
  -\omega - \frac{7\sigma^2}{96}x^2 + \frac{\sigma^2 g}{2x} = 0 .
\]
Adding them gives $x^3 = 12g/7$, and then
$\omega = \tfrac{7}{96}\sigma^2x^2 + \tfrac{7}{48}\sigma^2x^2
= \tfrac{7}{32}\sigma^2 x^2$. Hence $A = \tfrac{\sigma^2}{8}(1 + \tfrac74x^2)$ and
$s^2 = \sigma^2/(8A) = (1 + \tfrac74 x^2)^{-1}$. The neglected terms are smaller by a
factor $O(x)$, which changes $s_\star$ by $O(x^3) = O(g)$. Finally, for each $x$ the
profit is strictly decreasing in $\Lbar$, hence in $s$, so the maximal profit is
positive below $s_\star$ and negative above it.
\qed
\end{proof}

\begin{proof}[Proof of Theorem~\ref{thm:Vstar}]
On the active slice the stationary average-reward equation of the exact dynamics,
under the rule of Assumption~\ref{ass:boundary} with $\Gamma \to \infty$, reads
$\tfrac12\sD^2\Delta_d V^*(n) + r(n) - \gamma = 0$ for $|n| < w^*$, with $\gamma$ the
long-run gain and $\Delta_d$ the discrete Laplacian of Lemma~\ref{lem:occupation}. Every in-range offset covers the
active tick, so with locally flat volume $r(n) \equiv \Rcyc$, and the equation is
$\tfrac12\sD^2\Delta_d V^* = \gamma - \Rcyc$. For the quadratic~\eqref{eq:Vstar},
$\Delta_d n^2 = 2$ and $\Delta_d n = 0$ exactly on the integer lattice, so
$\tfrac12\sD^2\Delta_d V^* = -\sD^2(\varphi + 2g)/(2(w^*)^2) = -R_0(w^*)$, and the
equation holds with $\gamma = \Rcyc - R_0 = \Pi(w^*)$. Rebalancing resets the offset to $0$ at the cost of Definition~\ref{def:cost}, so
$V^*(w^*) = V^*(0) - \ILe|_{q'=b} - g$ and $V^*(-w^*) = V^*(0) - \ILe|_{q'=a} - g$.
The half-sum of these conditions is consistent with the curvature, since the sum of
the two IL values is $\varphi$, and their half-difference gives the linear
coefficient $-\eta/(2w^*)$. At the free-entry margin the maximal profit is
zero, so $\gamma = 0$. The equation and the two boundary conditions involve only
differences of $V^*$, so $C$ is free. With $V^*(\text{out}) = 0$, a provider out of
the pool does not gain by entering iff $C \le g$, and an active provider does not
gain by leaving at a rebalancing iff $C \ge 0$.
\qed
\end{proof}

\begin{proof}[Proof of Proposition~\ref{prop:offset}]
With $\Gamma \to \infty$ the forward equation of the offset on $|n| < w^*$ is the
walk killed at $\pm w^*$ with its flux re-injected at the origin,
\[
  \frac{d}{dt}\rho_t = \tfrac12\sD^2\Delta_d\rho_t + J_t\mathbf{1}\{n = 0\},
  \qquad
  J_t = \tfrac12\sD^2\bigl[\rho_t(w^*{-}1) + \rho_t(-w^*{+}1)\bigr],
\]
with $\rho_t(\pm w^*) = 0$.
Stationarity requires $\tfrac12\sD^2\Delta_d\rho = -J\,\mathbf{1}\{n = 0\}$, the
equation for the Green's function of Lemma~\ref{lem:occupation}, so
$\rho \propto G_{w^*} \propto w^* - |n|$, and normalization gives $\rho_{w^*}$. The reset walk is an irreducible Markov chain on the finite set
$\{-w^*+1,\ldots,w^*-1\}$, so its stationary law is unique. The flux is
$J = \tfrac12\sD^2\cdot 2/(w^*)^2 = \sD^2/(w^*)^2 = \leff$.
\qed
\end{proof}

\begin{proof}[Proof of Theorem~\ref{thm:MFE_triple}]
By Theorem~\ref{thm:Vstar}, $V^*$ is the value of the boundary rule at
width $w^*$ against $\Lbar$, with zero gain because the margin is $s_\star$. By
Proposition~\ref{prop:offset} the offset law is $\rho_{w^*}$, and every active
position covers the active tick, so the aggregate is~\eqref{eq:Lbar}. A provider
switching to another width $w'$ earns $\Pi(w') \le \Pi(w^*) = 0$, since $w^*$
maximizes $\Pi$ at this aggregate, and a provider staying out earns $0$. No
deviation within boundary-triggered provision is therefore profitable.
\qed
\end{proof}

\end{document}